\documentclass[a4paper]{article}

\makeatletter
\def\@seccntformat#1{\@ifundefined{#1@cntformat}%
   {\csname the#1\endcsname\quad}  
   {\csname #1@cntformat\endcsname}
}
\let\oldappendix\appendix 
\renewcommand\appendix{%
    \oldappendix
    \newcommand{\section@cntformat}{\appendixname~\thesection\quad}
}
\makeatother

\usepackage[pdftex]{graphicx}

\usepackage{cite}
\usepackage[normalem]{ulem}
\usepackage{comment}

\usepackage[pdftex]{xcolor}

\newcommand{\highlight}[1]{#1}
\usepackage{amssymb,amsfonts,amsmath}
\usepackage[top=25truemm,bottom=25truemm,left=20truemm,right=20truemm]{geometry}
\usepackage{bm}
\usepackage{cases}

\usepackage{lineno}
\newcommand*\patchAmsMathEnvironmentForLineno[1]{
  \expandafter\let\csname old#1\expandafter\endcsname\csname #1\endcsname
  \expandafter\let\csname oldend#1\expandafter\endcsname\csname end#1\endcsname
  \renewenvironment{#1}
     {\linenomath\csname old#1\endcsname}
     {\csname oldend#1\endcsname\endlinenomath}}
\newcommand*\patchBothAmsMathEnvironmentsForLineno[1]{
  \patchAmsMathEnvironmentForLineno{#1}
  \patchAmsMathEnvironmentForLineno{#1*}}
\AtBeginDocument{
\patchBothAmsMathEnvironmentsForLineno{equation}
\patchBothAmsMathEnvironmentsForLineno{align}
\patchBothAmsMathEnvironmentsForLineno{flalign}
\patchBothAmsMathEnvironmentsForLineno{alignat}
\patchBothAmsMathEnvironmentsForLineno{gather}
\patchBothAmsMathEnvironmentsForLineno{multline}
}

\usepackage[ruled,vlined]{algorithm2e}
\usepackage{setspace}	
\SetCommentSty{mycommfont}
\usepackage{multicol} 

\usepackage{setspace}

\usepackage{framed}	
\usepackage{lscape}	
\usepackage{calc}	

\newtheorem{theorem}{Theorem}

\newtheorem{corollary}{Corollary}
\newtheorem{definition}{Definition}
\newtheorem{lemma}{Lemma}

\title{Adapting paths against zero-determinant strategies in repeated prisoner's dilemma games}
\bigskip
\author{Daiki Miyagawa${}^{1*}$, Azumi Mamiya${}^{2}$ and Genki Ichinose${}^{1}$
\ \\
\ \\
${}^{1}$
Department of Mathematical and Systems Engineering, Shizuoka University, \\3-5-1 Johoku, Naka-ku, Hamamatsu, 432-8561, Japan\\
${}^{2}$
Nagoya Works, Mitsubishi Electric Corporation, \\5-1-14, Yada-minami, Higashi-ku, Nagoya,  461-8670, Japan\\
$^*$ Corresponding author (miyagawa.daiki.18@shizuoka.ac.jp)}

\begin{document}

\maketitle

\section*{Abstract}
Long-term cooperation, competition, or exploitation among individuals can be modeled through repeated games.
In repeated games, Press and Dyson discovered zero-determinant (ZD) strategies that enforce a special relationship between two players.
This special relationship implies that a ZD player can unilaterally impose a linear payoff relationship to the opponent regardless of the opponent's strategies.
A ZD player also has a property that can lead the opponent to an unconditional cooperation if the opponent tries to improve its payoff.
This property has been mathematically confirmed by Chen and Zinger.
Humans often underestimate a payoff obtained in the future.
However, such discounting was not considered in their analysis.
Here, we mathematically explored whether a ZD player can lead the opponent to an unconditional cooperation even if a discount factor is incorporated.
Specifically, we represented the expected payoff with a discount factor as the form of determinants and calculated whether the values obtained by partially differentiating each factor in the strategy vector become positive.
As a result, we proved that the strategy vector ends up as an unconditional cooperation even when starting from any initial strategy.
This result was confirmed through numerical calculations.
We extended the applicability of ZD strategies to real world problems.

\section*{Keywords}
Prisoner's dilemma, 
repeated games, 
discount factor, 
zero-determinant strategies, 
adaptive player, 
adapting path

\section{Introduction\label{sec:introduction}}
Societies consist of various interactions among individuals such as cooperation, competition, and exploitation.
Cooperative relationships are desirable for society because competition can be exhaustive.
However, there is always an incentive to exploit a cooperative relationship, which can be modeled using the prisoner's dilemma (PD) game.
Consider two players, in which each player selects one of two strategies: cooperation or defection.
Then, depending on the actions selected by the two players, the payoff is allocated to each.
A defection is the dominant strategy for both players because it yields a higher payoff than cooperation regardless of what strategy the opponent adopts.
Thus, in the one-shot PD, defection is a unique Nash equilibrium.
However, cooperation becomes a possible outcome in the repeated PD (RPD) game 
because it is beneficial for both players to build a cooperative relationship with each other when the game is repeated.
This mutual cooperation is called direct reciprocity \cite{Trivers1971QRevBiol,Nowak2006book,Sigmund2010book}.

Although the RPD game has been long studied, Press and Dyson suddenly discovered a completely new property, namely, the existence of zero-determinant (ZD) strategies.
They studied a one-to-one interaction in the RPD and showed that one of the two players can unilaterally enforce a linear payoff relationship on the opponent.
This characteristic has attracted significant attention from researchers in evolutionary games.
Thus, ZD strategies have been studied in the context of evolutionary games \cite{Akin2016ErgodicTheory, Adami2013NatComm, Hilbe2013PNAS, Hilbe2013PlosOne-zd, ChenZinger2014JTheorBiol, Szolnoki2014PhysRevE-zd, Szolnoki2014SciRep-zd, WuRong2014PhysRevE-zd, Hilbe2015JTheorBiol, LiuLi2015PhysicaA-zd, Xu2017PhysRevE-zd, Wang2019Chaos, Stewart2013PNAS, Mao2018EPL, Xu2019Neurocomputing}.
In addition to evolutionary games, ZD strategies have been studied from various directions.
Examples include games with observation errors \cite{Hao2015PhysRevE,MamiyaIchinose2019JTheorBiol}, multiplayer games \cite{Hilbe2014PNAS-zd,Hilbe2015JTheorBiol,Pan2015SciRep-zd,Milinski2016NatComm,Stewart2016PNAS, UedaTanaka2020PlosOne}, continuous action spaces \cite{Mcavoy2016PNAS,Milinski2016NatComm,Stewart2016PNAS,Mcavoy2017TheorPopulBiol}, alternating games \cite{Mcavoy2017TheorPopulBiol}, asymmetric games \cite{EngelFeigel2019ApplMathComput}, animal contests \cite{EngelFeigel2018PhysRevE}, human reactions to computerized ZD strategies \cite{Hilbe2014NatComm,Wang2016NatComm-zd}, and human-human experiments \cite{Hilbe2016PlosOne,Milinski2016NatComm, Becks2019NatComm}.

Among the various ZD strategies, of particular interest are the so-called Equalizer, Extortion \cite{Press2012PNAS}, and Generous strategies \cite{Stewart2013PNAS}.
Equalizer can fix the opponent's payoff \cite{Press2012PNAS}.
Extortion never results in a loss to any opponent in a one-to-one competition in terms of the expected payoffs.
However, Generous strategies always obtain lower payoffs than that of the opponent, with the exception of mutual cooperation \cite{Stewart2013PNAS}.
Extortion and Generous strategies are included as positively correlated ZD (pcZD) strategies \cite{ChenZinger2014JTheorBiol}.

In addition to a linear payoff relationship, ZD strategies are known to impose the following to the opponent.
That is, when the opponent player tries to improve its own payoff, the opponent is led in an unconditional cooperation \cite{Press2012PNAS}.
This property was predicted by Press and Dyson, 
and was numerically confirmed. 
Later, Chen and Zinger mathematically proved that this prediction is true in a subset of the RPD games \cite{ChenZinger2014JTheorBiol}.
However, this situation is the simplest case.
In reality, payoffs obtained in the future are underestimated. 
Thus, it is quite natural to incorporate a discount factor into the model.
Some studies have revealed the existence of ZD strategies with a discount factor \cite{Hilbe2015GamesEconBehav, Mcavoy2016PNAS, Mcavoy2017TheorPopulBiol, IchinoseMasuda2018JTheorBiol, GovaertCao2019arXiv}.
Hence, in this study, we incorporate a discount factor into the model and explore whether the opponent is led to an unconditional cooperation.
As a result, we first show that the expected payoffs between the two players are represented by the form of determinants. 
Then, using a similar technique as Chen and Zinger~\highlight{{\cite{ChenZinger2014JTheorBiol}}}, we mathematically show that this is true. 
\highlight{Our contribution is that we introduce the discount factor, which was ignored in the related work~\cite{ChenZinger2014JTheorBiol}, and derived the conditions to reach an unconditional cooperation under $0<\delta<1$. }

\section{Model}
We consider the symmetric two-player ($X$ and $Y$) RPD game. 
Player $i\in\{X,Y\}$ either cooperates (C) or defects (D) at each round. 
Such action is depicted as $a_i\in\{{\rm C},{\rm D}\}$. 
Then, $X$ ($Y$) obtains payoff $u_X$ ($u_Y$) depending on the set of actions between $X$ and $Y$ ($\bm a=(a_X,a_Y)$): 
$u_X(\bm a)=R$ and $u_Y(\bm a)=R$ for $\bm a=({\rm C},{\rm C})$; 
$u_X(\bm a)=S$ and $u_Y(\bm a)=T$ for $\bm a=({\rm C},{\rm D})$; 
$u_X(\bm a)=T$ and $u_Y(\bm a)=S$ for $\bm a=({\rm D},{\rm C})$; and 
$u_X(\bm a)=P$ and $u_Y(\bm a)=P$ for $\bm a=({\rm D},{\rm D})$. 
We can represent these payoffs through the following payoff matrix (row, $a_X$; column, $a_Y$): 
%
\begin{align}
\bordermatrix{
 &  {\rm C} &  {\rm D} \cr
{\rm C} & (R,R) & (S,T) \cr
{\rm D} & (T,S) & (P,P) \cr}. 
\label{eq:pMatrix_}	
\end{align}
The RPD game is defined by the following conditions:
\begin{align}
S<P<R<T,\ T+S<2R. 
\label{eq:PD_condS_}	
\end{align}

Here, we set $(R,P)=(1,0)$ to reduce the number of variables while keeping the relationship of Eq.~\eqref{eq:PD_condS_}.
Consequently, Eqs.~\eqref{eq:pMatrix_} and \eqref{eq:PD_condS_} are respectively modified into the following: 
%
\begin{align}
\bordermatrix{
 &  {\rm C} &  {\rm D} \cr
{\rm C} & (1,1) & (S,T) \cr
{\rm D} & (T,S) & (0,0) \cr}, \label{eq:pMatrix} \\
S<0,\ 1<T,\ T+S<2. \label{eq:PD_condS}	
\end{align}

\vskip 1em
We consider memory-1 strategies. 
Memory-1 strategies use the actions of both players at round $r-1$ to make their decisions at round $r$. 
We express a memory-1 strategy as a 5-tuple. 
The strategy of player $X$ is 
\begin{align}
\bm p=(p_0; p_1, p_2, p_3, p_4), 
\label{eq:pvector}	
\end{align}
where $0\le p_j\le1\ (j\in\{0,1,2,3,4\})$. 
In addition, $p_1$ is the cooperation probability for the current round when $\bm a=({\rm C},{\rm C})$ at the last round; 
$p_2$, $p_3$, and $p_4$ correspond to (C,D), (D,C), and (D,D), respectively; 
and $p_0$ is the cooperation probability at the first round. 
%
%
As with player $X$, we define the strategy of player $Y$ as below:
\begin{align}
\bm q=(q_0; q_1, q_2, q_3, q_4), 
\label{eq:qvector}	
\end{align}
where $0\le q_j\le1\ (j\in\{0,1,2,3,4\})$. 
In addition, $q_1,\ q_2,\ q_3$, and $q_4$ correspond to $\bm a=({\rm C},{\rm C})$, (D,C), (C,D), and (D,D), respectively, 
and $q_0$ is the cooperation probability at the first round. 

\vskip 1em
We define the stochastic state of two players at round $r$ as $\bm v(r)=(v_1(r), v_2(r), v_3(r), v_4(r))$, 
where $v_1$, $v_2$, $v_3$, and $v_4$ correspond to $\bm a=({\rm C},{\rm C})$, (C,D), (D,C), and (D,D), respectively. 
Using this, we derive the expected payoff at round $r$ for player $i$ as $\bm v(r) \bm S_i$, 
where $\bm S_i$ is the payoff vector, such that $\bm S_X=(1,S,T,0)^T$ and $\bm S_Y=(1,T,S,0)^T$. 
Defining $s_i$ as the average payoff per game, we obtain
%
\begin{align}
s_i = (1-\delta) \sum^{\infty}_{r=0} \delta^r \bm v(r) \bm S_i, 
\label{eq:aveExp}	
\end{align}
where $\delta\in(0,1)$ is the discount factor, which means that the payoffs obtained in the future are underestimated. 
\highlight{Because this factor relatively increases the value of payoffs obtained in the first round, we must use $p_0$ and $q_0$ to determine the first game outcome even though the related study~\cite{ChenZinger2014JTheorBiol} did not consider it. }

The initial stochastic state is $\bm v(0)=(p_0 q_0, p_0 (1-q_0), (1-p_0) q_0, (1-p_0) (1-q_0))$. 
The transition matrix $M$ from the last round to the current round is given by the following:
%
\begin{align}
M=\left(
  \begin{array}{rrrr}
   p_1 q_1  &  p_1 (1-q_1)  &  (1-p_1) q_1  &  (1-p_1) (1-q_1)  \\
   p_2 q_3  &  p_2 (1-q_3)  &  (1-p_2) q_3  &  (1-p_2) (1-q_3)  \\
   p_3 q_2  &  p_3 (1-q_2)  &  (1-p_3) q_2  &  (1-p_3) (1-q_2)  \\
   p_4 q_4  &  p_4 (1-q_4)  &  (1-p_4) q_4  &  (1-p_4) (1-q_4)
  \end{array}
 \right).
\label{eq:transMatrix}	
\end{align}
We then obtain $\bm v(r)=\bm v(0) M^r$ at round $r$. 
Thus, substituting it for $\bm v(r)$ in Eq.~(\ref{eq:aveExp}), we obtain 
%
\begin{align}
s_i &= (1-\delta) \bm v(0) \sum^{\infty}_{r=0} (\delta M)^r \bm S_i \nonumber \\
    &= (1-\delta) \bm v(0) (I-\delta M)^{-1} \bm S_i, 
\label{eq:aveExpM}	
\end{align}
where $I$ is the $4\times4$ identity matrix. 
\highlight{
Note that the conditions under which the matrix $I-\delta M$ is inversible are $0\le p_j,q_j\le1$ and $0<\delta<1$ (described in detail in Appendix~\ref{app:inversible}). 
}
Define the mean distribution of $\bm v(r)$ as $\bm v^T \equiv (1-\delta) \bm v(0) (I-\delta M)^{-1}$. 
In \cite{MamiyaIchinose2021JTheorBiol}, the authors showed $\bm v(0)=\bm v^T M_0$, where $M_0$ is below: 
%
\begin{align}
M_0=\left(
  \begin{array}{rrrr}
   p_0 q_0  &  p_0 (1-q_0)  &  (1-p_0) q_0  &  (1-p_0) (1-q_0)  \\
   p_0 q_0  &  p_0 (1-q_0)  &  (1-p_0) q_0  &  (1-p_0) (1-q_0)  \\
   p_0 q_0  &  p_0 (1-q_0)  &  (1-p_0) q_0  &  (1-p_0) (1-q_0)  \\
   p_0 q_0  &  p_0 (1-q_0)  &  (1-p_0) q_0  &  (1-p_0) (1-q_0)
  \end{array}
 \right).
\label{eq:matrix0}	
\end{align}
We then obtain $\bm v^T = (1-\delta) \bm v^T M_0 (I-\delta M)^{-1}$. 
Multiplying both sides of the equation by $I-\delta M$ from the right, we obtain $\bm v^T (I-\delta M) = (1-\delta) \bm v^T M_0$. 
Therefore, we obtain $\bm v^T M'=\bm 0$, where $M'\equiv \delta M+(1-\delta)M_0 -I$. 
\highlight{
Because this equation is the same style as Eq.~(1) in \cite{Press2012PNAS}, we can write out $s_i$ in the following way on the basis of \cite{Press2012PNAS}: 
Since we have ${\rm Adj}(M')M' = O$ by applying Cramer's rule, where ${\rm Adj}(M')$ is the adjoint matrix of $M'$ and $O$ is $4\times4$ matrix all of which entries are zero, every row of ${\rm Adj}(M')$ is proportional to $\bm v^T$. 
We thus obtain $\bm v=\frac{\bm u}{\bm u\cdot\bm 1_4}$ because $v_1+v_2+v_3+v_4=1$ where $\bm u$ is the fourth row of ${\rm Adj}(M')$ and $\bm 1_4=(1,1,1,1)^T$. 
Now, we can derive the inner product between $\bm u$ and the arbitrary vector $\bm f=(f_1, f_2, f_3, f_4)^T$ as follows: 
}
%
\begin{align}
 \bm u\cdot\bm f = \det
 \left(
  \begin{array}{rrrr}
   -1 +\delta p_1 q_1 +(1 -\delta) p_0 q_0   &   -1 +\delta p_1 +(1 -\delta) p_0   &   -1 +\delta q_1 +(1 -\delta) q_0   &   f_1  \\
       \delta p_2 q_3 +(1 -\delta) p_0 q_0   &   -1 +\delta p_2 +(1 -\delta) p_0   &       \delta q_3 +(1 -\delta) q_0   &   f_3  \\
       \delta p_3 q_2 +(1 -\delta) p_0 q_0   &       \delta p_3 +(1 -\delta) p_0   &   -1 +\delta q_2 +(1 -\delta) q_0   &   f_2  \\
       \delta p_4 q_4 +(1 -\delta) p_0 q_0   &       \delta p_4 +(1 -\delta) p_0   &       \delta q_4 +(1 -\delta) q_0   &   f_4
  \end{array}
 \right). 
 \label{eq:determinant__}	
\end{align}
From Eq.~(\ref{eq:determinant__}), we represent the average expected payoff $s_i$ as follows: 
%
\begin{align}
s_i = \bm v\cdot\bm S_i=\frac{\bm u\cdot\bm S_i}{\bm u\cdot\bm 1_4}. 
\end{align}
%
Here, because we focus on player $Y$'s strategy $\bm q$, we define $D(\bm p, \bm q, \bm f)$ by exchanging the second row with the third row of Eq.~(\ref{eq:determinant__}), as shown below: 
\begin{align}
 D(\bm p,\bm q,\bm f) &= -\bm u\cdot\bm f \nonumber \\ 
 &= \det
 \left(
  \begin{array}{rrrr}
   -1 +\delta p_1 q_1 +(1 -\delta) p_0 q_0   &   -1 +\delta p_1 +(1 -\delta) p_0   &   -1 +\delta q_1 +(1 -\delta) q_0   &   f_1  \\
       \delta p_3 q_2 +(1 -\delta) p_0 q_0   &       \delta p_3 +(1 -\delta) p_0   &   -1 +\delta q_2 +(1 -\delta) q_0   &   f_2  \\
       \delta p_2 q_3 +(1 -\delta) p_0 q_0   &   -1 +\delta p_2 +(1 -\delta) p_0   &       \delta q_3 +(1 -\delta) q_0   &   f_3  \\
       \delta p_4 q_4 +(1 -\delta) p_0 q_0   &       \delta p_4 +(1 -\delta) p_0   &       \delta q_4 +(1 -\delta) q_0   &   f_4
  \end{array}
 \right). 
 \label{eq:determinant_}	
\end{align}
Thus, we obtain 
\begin{align}
s_i = \frac{D(\bm p,\bm q,\bm S_i)}{D(\bm p,\bm q,\bm 1_4)}. 
\label{eq:aveExpFinal}
\end{align}

If player $X$'s strategy $\bm p$ is a zero-determinant strategy, $\alpha s_X+\beta s_Y+\gamma=0$ is satisfied for any $\bm q$ in player $Y$. 
Therefore, we obtain the following four equations as ZD strategies: 
%
\begin{align}
   -1 +\delta p_1 +(1 -\delta) p_0   &=  \alpha+\beta+\gamma,		\label{eq:zd1}\\
   -1 +\delta p_2 +(1 -\delta) p_0   &=  \alpha S+\beta T+\gamma,	\label{eq:zd2}\\
       \delta p_3 +(1 -\delta) p_0   &=  \alpha T+\beta S+\gamma,	\label{eq:zd3}\\
       \delta p_4 +(1 -\delta) p_0   &=  \gamma.					\label{eq:zd4}
\end{align}
Calculating $(\ref{eq:zd2})+(\ref{eq:zd3})-2\times(\ref{eq:zd4})-((\ref{eq:zd1})-(\ref{eq:zd4}))(T+S)$, we obtain the relational expression as follows: 
%
\begin{align}
\delta p_2+\delta p_3=1+2\delta p_4-(1-\delta p_1+\delta p_4)(T+S). 
 \label{eq:lincond}
\end{align}

If $\alpha\ne0$, we can transform Eqs.~(\ref{eq:zd1})–(\ref{eq:zd4}) by substituting $\alpha=\phi$, $\beta=-\phi\chi$, and $\gamma=\phi(\chi-1)\kappa$ as below:
%
%
%
\begin{align}
   \delta p_1 +(1 -\delta) p_0   &= 1 -\phi(\chi-1)(1-\kappa),			\label{eq:zd5}\\
   \delta p_2 +(1 -\delta) p_0   &= 1 -\phi(\chi T-S-(\chi-1)\kappa),	\label{eq:zd6}\\
   \delta p_3 +(1 -\delta) p_0   &=    \phi(T-\chi S+(\chi-1)\kappa),	\label{eq:zd7}\\
   \delta p_4 +(1 -\delta) p_0   &=    \phi(\chi-1)\kappa.				\label{eq:zd8}
\end{align}
Now, we can rewrite $\alpha s_X+\beta s_Y+\gamma=0$ using $\chi$ and $\kappa$ as below: 
\begin{align}
s_X-\kappa = \chi(s_Y-\kappa). 
\end{align}

If player $X$ is a positively correlated ZD strategy ($\chi\ge1$), we obtain the following relational expressions by comparing Eqs.~(\ref{eq:zd5}) and (\ref{eq:zd6}), and Eqs.~(\ref{eq:zd7}) and (\ref{eq:zd8}), respectively, as follows: 
%
\begin{align}
p_1>p_2,p_3>p_4,  \nonumber \\
\therefore p_1,\hat{p_2},p_3,\hat{p_4}>0, 
\label{eq:pcZDcond}
\end{align}
where $\hat{x}\equiv 1-x$. 
%
In \cite{MamiyaIchinose2021JTheorBiol}, the authors show that the smaller $\delta$ is, the narrower the existence range of the pcZD strategy. 
Therefore, we restrict the range of $\delta$ as $\delta_c < \delta < 1$, 
where $\delta_c$ is the largest $\delta$ at which the pcZD strategy cannot exist. 
According to Eq.~(27) in \cite{MamiyaIchinose2021JTheorBiol}, $\delta_c$ is given by
\begin{align}
\delta_c = \max\left(\frac{T-1}{T},\frac{-S}{1-S}\right)>0. 
\label{eq:delta_c}
\end{align}


\highlight{From then on, the restriction of $\delta_c<\delta<1$ is imposed on $\delta$ when pcZD is assumed.}

\section{Result}
\begin{figure}[hbtp]
\centering 
\includegraphics[width=\textwidth-10mm]{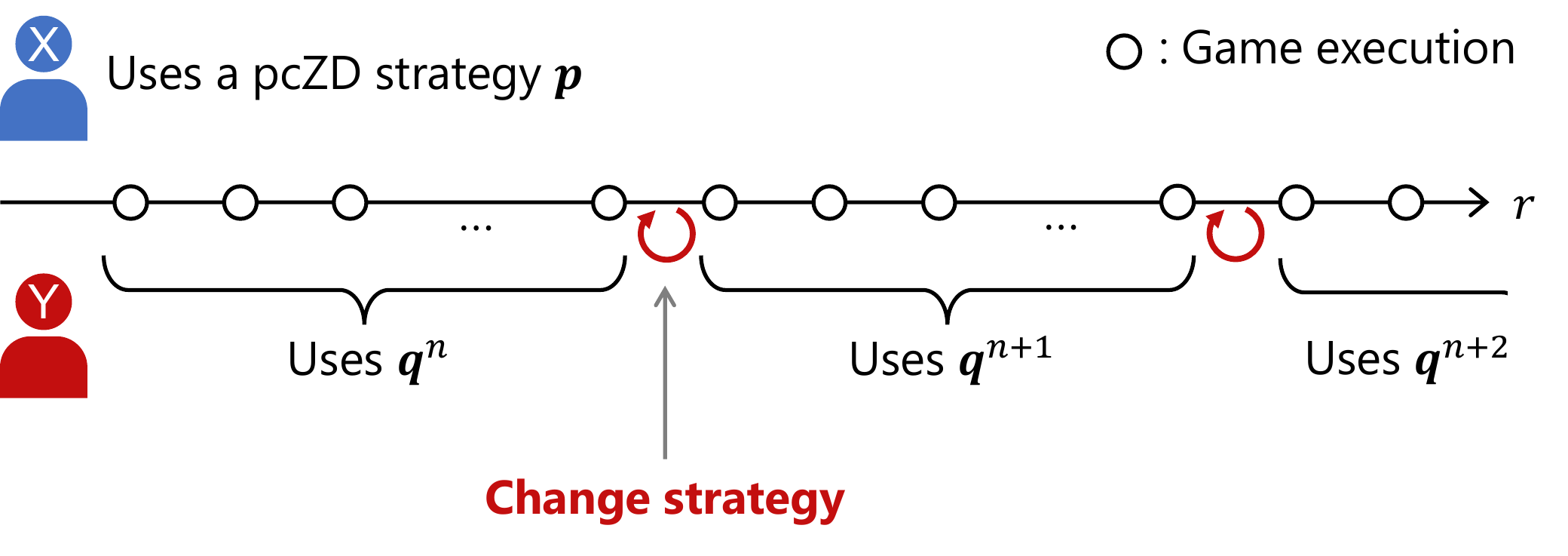}
\caption{Flow of games and strategy changes. Players $X$ (blue) and $Y$ (red) play games (white circles) repeatedly. While player $X$ keeps using a pcZD strategy, player $Y$ can change its strategy (rounded red arrow). }
\label{fig:flow}
\end{figure}

We focus on the opponent facing a ZD player. 
The opponent tries to improve its payoff by changing its strategy, 
which we call an \textit{adaptive player}. 
%
\highlight{
The adaptive player uses the average expected payoff $s_Y$ to change from its current strategy $\bm q^{n}$ to a new strategy $\bm q^{n+1}$. 
In other words, it choose $\bm q^{n+1}$ that will give it a larger $s_Y$ than it can obtain by continuing to use $\bm q^{n}$. 
Figure ~\ref{fig:flow} shows the flow of executing games between players and changing $Y$'s strategy. 
}

Moreover, \highlight{based on the definition of \cite{ChenZinger2014JTheorBiol}}, we also define an \textit{adapting path} as the locus in which the adaptive player changes its strategy from the initial strategy until it can no longer improve its payoff:

\begin{definition}
\begin{framed}

\highlight{
An adapting path of an adaptive player is the discrete map $\lambda : A\rightarrow[0,1]^5\ (A=\{0,1,2,\cdots,N\},\ N\in\mathbb{N})$, such that 

\vspace{4mm}
\noindent
(A1) $s_Y(\lambda(n_1)) < s_Y(\lambda(n_2))\ (0\le n_1<n_2 \le N)$;

\noindent
(A2) There is no discrete map $\tilde{\lambda} : \{0,1,2,\cdots,N,\cdots,\tilde{N}\}\rightarrow[0,1]^5$ where $\forall n\in A\ \lambda(n)=\tilde{\lambda}(n)$ satisfying (A1). 
}
\label{def:adapting_path}	
\end{framed}
\end{definition}
We interpret $n$ and $\lambda$ as a time variable and strategy vector, respectively. 
Then, an adapting path $\lambda(A)$ depicts the locus of the strategy vector defined from $n=0$ to $N$. 
(A1) indicates that the function $s_Y$ (payoff variable) increases as time passes. 
(A2) shows that the adapting path cannot extend further beyond the last time $N$. 
Thus, the locus, which ends at a point which it can still improve $s_Y$, is not an adapting path.

Here, we prove the following theorem for all adapting paths for a case in which the adaptive player faces the pcZD strategies:

\begin{theorem}
\begin{framed}
Assume that player $X$ uses a pcZD strategy and player $Y$ is the adaptive player facing with $X$. 
For the RPD games with discount factor $\delta\in(\delta_c,1)$, we use the following payoff relationships: 
%
\begin{align}
S<0,\ 1<T,\ \ 0<T+S<2, 
\label{eq:PD_cond}
\end{align}
which have more limited conditions than the ordinary RPD games owing to $0<T+S$. 

Under this assumption, all adapting paths with any initial $\bm q$ lead $Y$ to a strategy with $(q_0, q_1, q_2) = (1, 1, 1)$ if $p_0<1$ or $p_1<1$, or with $(q_0, q_1) = (1, 1)$ if $(p_0, p_1) = (1, 1)$.
\label{theor:main}	
\end{framed}
\end{theorem}

Although this theorem is similar to Theorem 1 proven in \cite{ChenZinger2014JTheorBiol}, the conditions of $p_0$ and $q_0$ are novel.

We describe the outline of our proof.
First, we calculate the partial derivatives of $s_Y$ for $q_j\ (j\in\{0,1,2,3,4\})$ and check their signs to know how to improve $s_Y$.
Instead of using $s_Y$, we actually use $\partial s_X/\partial q_j$ to simplify the equation because the sign of $\partial s_X/\partial q_j$ is equal to that of $\partial s_Y/\partial q_j$ owing to the positive linear relationship between $s_Y$ and $s_X$. 

The important point proving this main theorem is expressing $s_X$ as a form of determinants which include discount factor $\delta$ (shown in Eq.~(\ref{eq:aveExpFinal})). 
This expression allows us to factorize $\partial s_X/\partial q_j$ using the method in \cite{ChenZinger2014JTheorBiol}. 
Thus, it is sufficient to examine the corner cases of $\bm q$ to check the sign of each factor because each factor is linear to some elements of $q_j$. 

As the result of calculations, we found that (1) $\partial s_Y/\partial q_{\ell}\ge0\ (\ell\in\{1,2,3,4\})$ for every $\bm q$, whereas (2) $\partial s_Y/\partial q_0$ occasionally becomes less than 0. 
Through (1), $q_{\ell}$ can increase regardless of $q_0$ until every $q_{\ell}$ reaches a value of 1 or its partial derivative reaches zero. 
We regard this state as a \textit{relay state}, at which we found that $\partial s_Y/\partial q_0>0$. 
From this relay state, $q_0$ also increases, then finally stops at strategies that satisfy $(q_0,q_1,q_2)=(1,1,1)$ or $(p_0,p_1,q_0,q_1)=(1,1,1,1)$.
These strategies correspond to the final state of all adapting paths.

\section{Proof}
To know where the destination of every adapting path is, 
we must derive how $\bm q$ should be changed to improve $s_Y$. 
It is then effective to examine the signs of partial derivatives of $s_Y$ for $q_j\ (j\in\{0,1,2,3,4\})$. 
We use $\partial s_X/\partial q_j$ because we can factorize it into a few simple factors. 
Note that the sign of $\partial s_X/\partial q_j$ is equal to that of $\partial s_Y/\partial q_j$ ($\partial s_Y/\partial q_j \propto \partial s_X/\partial q_j$) 
because there is a positively correlation between $s_X$ and $s_Y$ owing to a pcZD strategy of player $X$. 

Using the quotient rule, the equation of partial derivatives of $s_X$ is transformed into the following form: 
%
\begin{align}
 \frac{\partial s_X}{\partial q_j} &= \frac{\partial}{\partial q_j}\left(\frac{D(\bm p,\bm q,\bm S_X)}{D(\bm p,\bm q,\bm 1_4)}\right) \nonumber \\
 &= \frac
		{\displaystyle{D(\bm p, \bm q, \bm 1_4)\cdot \frac{\partial D(\bm p, \bm q, \bm S_X)}{\partial q_j}
   -\frac{\partial D(\bm p, \bm q, \bm 1_4)}{\partial q_j}\cdot D(\bm p, \bm q, \bm S_X)}}
		{D(\bm p,\bm q,\bm 1_4)^2}; \nonumber \\
\nonumber \\
 \therefore D(\bm p, \bm q, \bm 1_4)^2 \frac{\partial s_X}{\partial q_j} 
 &= D(\bm p, \bm q, \bm 1_4)\cdot \frac{\partial D(\bm p, \bm q, \bm S_X)}{\partial q_j}
   -\frac{\partial D(\bm p, \bm q, \bm 1_4)}{\partial q_j}\cdot D(\bm p, \bm q, \bm S_X) \nonumber \\
 &= \det \left(
  \begin{array}{cc}
   D(\bm p, \bm q, \bm 1_4)  &  D(\bm p, \bm q, \bm S_X)  \\
   \\
   \displaystyle \frac{\partial D(\bm p, \bm q, \bm 1_4)}{\partial q_j}  &  \displaystyle \frac{\partial D(\bm p, \bm q, \bm S_X)}{\partial q_j}
  \end{array}
 \right). 
 \label{eq:partial}
\end{align}
Based on Eq.~(\ref{eq:partial}), to obtain the sign of $\partial s_X/\partial q_j$, we found that we must verify the existence of singularities of $s_X$, and then examine the sign on the right-hand side (RHS) of Eq.~(\ref{eq:partial}). 

We factorized the RHS of Eq.~(\ref{eq:partial}) for simplification. 
For $\ell\in\{1,2,3,4\}$, the RHS is factorized as below using the transformation proposed in \cite{ChenZinger2014JTheorBiol}: 
%
\highlight{
\begin{align}
 D(\bm p, \bm q, \bm 1_4)^2 \frac{\partial s_X}{\partial q_{\ell}} 
 &= \delta (1-\delta p_2-(1-\delta p_1)S +\delta p_4 (1 -S)) \left<M_{\ell}\right> \mathfrak{d}_{\ell}, 
 \label{eq:factorized}
\end{align}
}
where $\left<M_{\ell}\right>$ is the determinant eliminated the $\ell$th row and the fourth column from $D(\bm p, \bm q, \bm f)$
and $\mathfrak{d}_{\ell}$ is also a determinant (we show the derivation of $\mathfrak{d}_{\ell}$ in Appendix~\ref{app:factorization}). 

However, for $q_0$, the RHS is factorized as follows: 
\highlight{
\begin{align}
 D(\bm p, \bm q, \bm 1_4) \frac{\partial s_X}{\partial q_0} 
 &= (1 -\delta) (1-\delta p_2-(1-\delta p_1)S +\delta p_4 (1 -S)) \mathfrak{d}_0, 
 \label{eq:factorized_0}
\end{align}
}
where $\mathfrak{d}_0$ is also the determinant (shown in Appendix~\ref{app:factorization}). 

%
\vskip 1em
Initially, we confirm the existence of singularities of $s_X$ by examining $D(\bm p,\bm q,\bm 1_4)=0$. 
We apply the following operations to $D(\bm p, \bm q, \bm 1_4)$ for simplification: 
Subtract the fourth column multiplied by $(1-\delta)p_0 q_0$ from the first column, subtract the fourth column multiplied by $(1-\delta)p_0$ from the second column, subtract the fourth column multiplied by $(1-\delta)q_0$ from the third column. 
We then obtain the following: 
\begin{align}
 D(\bm p, \bm q, \bm 1_4) = \det
 \left(
  \begin{array}{rrrr}
   -1 +\delta p_1 q_1   &   -1 +\delta p_1   &   -1 +\delta q_1   &   1  \\
       \delta p_3 q_2   &       \delta p_3   &   -1 +\delta q_2   &   1  \\
       \delta p_2 q_3   &   -1 +\delta p_2   &       \delta q_3   &   1  \\
       \delta p_4 q_4   &       \delta p_4   &       \delta q_4   &   1
  \end{array}
 \right). \label{eq:Dpq1}
\end{align}
Equation~(\ref{eq:Dpq1}) shows that $D(\bm p, \bm q, \bm 1_4)$ is linear to $q_{\ell}$ but independent of $q_0$. 
Therefore, we can capture the range of $D(\bm p,\bm q,\bm 1_4)$ by checking its sign for all corner cases of $(q_1, q_2, q_3, q_4)$. 
We obtain Table~\ref{tab:Dpq1} by substituting each corner case (from $(0,0,0,0)$ to $(1,1,1,1)$) for $(q_1,q_2,q_3,q_4)$ in $D(\bm p,\bm q,\bm 1_4)$. 
%
\begin{table}[btp]
\centering
\caption{Values of the determinant $D(\bm p, \bm q, \bm 1_4)$ at the corners of $[0,1]^4$; $\hat{x} \equiv 1 -x$, $\dot{p_j} \equiv 1 -\delta p_j$, $\ddot{p_j} \equiv 1 -\delta^2 p_j$.} 
\scalebox{0.85}[0.85]{
\begin{tabular}{lllll}
\hline
$(q_1,q_2)$ & $(q_3,q_4)$ &       &       &       \\ \cline{2-5} 
            & (0,0)       & (0,1) & (1,0) & (1,1) \\ \hline
\\
\begin{tabular}{l} (0,0) \\ \\ \end{tabular} & 
\begin{tabular}{l} $\dot{p_2} +\delta p_4$ \\ \\ \end{tabular} &	
\begin{tabular}{l} $\delta^2 p_1 p_4 +(1+\delta) \dot{p_2}$ \\ $+\delta^2 p_3 \hat{p_4}$ \\ \end{tabular}   &	
\begin{tabular}{l} $\ddot{p_1} p_2 +\hat{p_2} \ddot{p_3}$ \\ $+\delta (1+\delta) p_4$ \\ \end{tabular}   &	
\begin{tabular}{l} $(1 +\delta) (1 -\delta^2 (p_1 -p_3) (p_2 -p_4)) $ \\ \\ \end{tabular}	
\\ \\
\begin{tabular}{l} (0,1) \\ \\ \end{tabular} &
\begin{tabular}{l} $(\dot{p_2} +\delta p_4) (\delta p_3 +\hat{\delta})$ \\ \\ \end{tabular}   &	
\begin{tabular}{l} $\delta^3 p_1 p_3$ \\ $+\dot{p_2} (\ddot{p_4} +\delta (1 +\delta) p_3)$ \\ $+\delta^2 \hat{\delta} p_1 p_4$ \end{tabular}   &	
\begin{tabular}{l} $\delta \ddot{p_1} p_3$ \\ $+\delta (\ddot{p_2} +\delta (1+\delta) p_3) p_4$ \\ $+\hat{\delta} (1 -\delta^2 p_1 p_2)$ \end{tabular}   &	
\begin{tabular}{l} $\hat{\delta} (1+\delta) +\delta^2 p_1 \hat{p_2}$ \\ $+\delta^2 \hat{p_1} \hat{p_4} +\delta (1+\delta) p_3$ \\ \end{tabular}	
\\ \\
\begin{tabular}{l} (1,0) \\ \\ \end{tabular} &
\begin{tabular}{l} $\dot{p_1} (\dot{p_2} +\delta p_4)$ \\ \\ \end{tabular} &	
\begin{tabular}{l} $\dot{p_1} \{ (1+\delta) \dot{p_2} +\delta^2 p_3 \}$ \\ $+\delta^2 (\delta \hat{p_2} +\hat{\delta} \hat{p_3}) p_4$ \\ \end{tabular} &	
\begin{tabular}{l} $\dot{p_1} (\ddot{p_3} +\delta(1 +\delta) p_4)$ \\ $+\delta^3 p_2 p_4$ \\ $+\delta^2 \hat{\delta} p_2 p_3$ \end{tabular} &	
\begin{tabular}{l} $(1 +\delta) \dot{p_1}$ \\ $+\delta^2 p_2 p_3 +\delta^2 \hat{p_3} p_4$ \\ \end{tabular}	
\\ \\
\begin{tabular}{l} (1,1) \\ \\ \end{tabular} &
\begin{tabular}{l} $\hat{\delta} (\dot{p_2} +\delta p_4) (\dot{p_1} +\delta p_3)$ \\ \\ \end{tabular} &	
\begin{tabular}{l} $(\dot{p_1} +\delta p_3) \dot{p_2} $\\ \\ \end{tabular} &	
\begin{tabular}{l} $(\dot{p_1} +\delta p_3) (\delta p_4 +\hat{\delta})$ \\ \\ \end{tabular} &	
\begin{tabular}{l} $\dot{p_1} +\delta p_3$ \\ \\ \end{tabular}	
\\ \hline
\end{tabular}
} 
\label{tab:Dpq1}
\end{table}
Every expression of Table~\ref{tab:Dpq1} is composed of some terms that contain $\delta$ and $\bm p$, 
where 
\begin{align}
\dot{p_j}\equiv1-\delta p_j,\ \ddot{p_j}\equiv1-\delta^2p_j. 
\label{eq:expressions}
\end{align}
Considering $0<\delta<1$, Eq.~(\ref{eq:expressions}) satisfies the following condition: 
\begin{align}
 \hat{\delta},\ \dot{p_j},\ \ddot{p_j} >0,\ j\in\{0,1,2,3,4\}. 
 \label{eq:delta_inequality}
\end{align}

Adapting Eq.~(\ref{eq:delta_inequality}) to Table~\ref{tab:Dpq1}, we found that every corner case is greater than zero. 
Therefore, we obtain the following lemma: 

%
\begin{lemma}
\begin{oframed}
For arbitrary $\bm p$ with discount factor $\delta\ (0<\delta<1)$, 
\begin{align}
 D(\bm p, \bm q, \bm 1_4) > 0
\end{align}
for all $\bm q\in[0,1]^5$. 
\label{lemma:Dpq1}
\end{oframed}
\end{lemma}

Furthermore, we obtain the following corollary: 
\begin{corollary}
\begin{oframed}
For arbitrary $\bm p$ with discount factor $\delta\ (0<\delta<1)$, player $Y$'s average expected payoff $s_Y$ is partially differentiable over the whole range of $\bm q\in[0,1]^5$, and thus has no singularities. 
\label{corollary:singular}
\end{oframed}
\end{corollary}

%
\vskip 1em
Second, we checked the sign of $\partial s_X/\partial q_j$ by examining that of the RHS in Eqs.~(\ref{eq:factorized}) and (\ref{eq:factorized_0}). 

Focussing on each factor of Eq.~(\ref{eq:factorized}), it is obvious that the first factor is greater than zero because $0<\delta<1$. 
The first factor of Eq.~(\ref{eq:factorized_0}) is also greater than zero. 
Moreover, the second factor \highlight{$1-\delta p_2-(1-\delta p_1)S+\delta p_4(1-S)$} (which is common between Eqs.~(\ref{eq:factorized}) and (\ref{eq:factorized_0})) is also greater than zero because of $\bm p\in[0,1]^5$, as shown in Eqs.(\ref{eq:PD_cond}), and Eq.~(\ref{eq:delta_inequality}). 

Next, we examined the sign of $(-1)^{\ell}\left<M_{\ell}\right>\ (\ell\in\{1,2,3,4\})$. 
Determinant $\left<M_{\ell}\right>$ is independent of $q_{\ell}$ but linear to the others. 
Thus, we must examine the corner cases of $\bm q_{-\ell}\equiv\{q_j; j\in\{0,1,2,3,4\}, j\ne\ell\}$. 
Table~\ref{tab:M_l} shows all corner cases of $(-1)^{\ell}\left<M_{\ell}\right>$. 
%
\begin{table}[btp]
\centering
\caption{Values of the determinants $\left<M_\ell\right>$ at the corners of $[0,1]^4$ for $\bm q_{-\ell}$; $\hat{x}\equiv1-x$, $\dot{p_j} \equiv 1 -\delta p_j$, $\ddot{p_j} \equiv 1 -\delta^2 p_j$. }
\scalebox{0.8}[0.8]{
\begin{tabular}{lllll}
\hline
$\bm q_{-\ell}$   &   $-\left<M_1\right>$   &   $\left<M_2\right>$   &   $-\left<M_3\right>$   &   $\left<M_4\right>$ \\ \hline

\begin{tabular}{l} (0,0,0,0) \\ \\ \end{tabular} & 
\begin{tabular}{l} 0         \\ \\ \end{tabular} &
\begin{tabular}{l} 0         \\ \\ \end{tabular} &
\begin{tabular}{l} $\delta p_4 +\hat{\delta} p_0$ \\ \\ \end{tabular} &
\begin{tabular}{l} $\delta \hat{p_2} +\hat{\delta} \hat{p_0}$ \\ \\ \end{tabular}
\\

\begin{tabular}{l} (0,0,0,1) \\ \\ \end{tabular} & 
\begin{tabular}{l} $\delta p_4 (\delta \hat{p_2} +\hat{\delta} \hat{p_0})$ \\ \\ \end{tabular} &
\begin{tabular}{l} $\delta \hat{p_4} (\delta \hat{p_2} +\hat{\delta} \hat{p_0})$ \\ \\ \end{tabular} &
\begin{tabular}{l} $\delta^2 (p_1 p_4 +p_3 \hat{p_4})$ \\ $+\hat{\delta^2} p_0$ \\ \end{tabular} &
\begin{tabular}{l} $\delta^2 (\hat{p_1} p_2 +\hat{p_2} \hat{p_3})$ \\ $+\hat{\delta^2} \hat{p_0}$ \\ \end{tabular}
\\ \\

\begin{tabular}{l} (0,0,1,0) \\ \\ \end{tabular} & 
\begin{tabular}{l} $\delta p_2 (\delta p_4 +\hat{\delta} p_0)$ \\ \\ \end{tabular} &
\begin{tabular}{l} $\delta \hat{p_2} (\delta p_4 +\hat{\delta} p_0)$ \\ \\ \end{tabular} &
\begin{tabular}{l} $(\delta p_3 +\hat{\delta}) (\delta p_4 +\hat{\delta} p_0)$ \\ \\ \end{tabular} &
\begin{tabular}{l} $(\delta p_3 +\hat{\delta}) (\delta \hat{p_2} +\hat{\delta} \hat{p_0})$ \\ \\ \end{tabular}
\\

\begin{tabular}{l} (0,0,1,1) \\ \\ \end{tabular} & 
\begin{tabular}{l} $\delta^3 (p_2 p_3 +\hat{p_3} p_4)$ \\ $+\delta \hat{\delta^2} (p_0 p_2 +\hat{p_0} p_4)$ \\ \end{tabular} &
\begin{tabular}{l} $\delta^3 (p_1 \hat{p_2} +\hat{p_1} \hat{p_4})$ \\ $+\delta \hat{\delta^2} (p_0 \hat{p_2} +\hat{p_0} \hat{p_4})$ \\ \end{tabular} &
\begin{tabular}{l} $\delta^2 p_1 (\delta p_3 +\hat{\delta} p_4)$ \\ $+p_0 (\hat{\delta} \ddot{p_4} +\delta \hat{\delta^2} p_3)$ \\ \end{tabular} &
\begin{tabular}{l} $\delta^2 \hat{p_1} (\delta p_3 +\hat{\delta} p_2)$ \\ $+\hat{p_0} (\hat{\delta} \ddot{p_2} +\delta \hat{\delta^2} p_3)$  \\ \end{tabular}
\\ \\

\begin{tabular}{l} (0,1,0,0) \\ \\ \end{tabular} & 
\begin{tabular}{l} 0         \\ \\ \end{tabular} &
\begin{tabular}{l} 0         \\ \\ \end{tabular} &
\begin{tabular}{l} $\dot{p_1} (\delta p_4 +\hat{\delta} p_0)$ \\ \\ \end{tabular} &
\begin{tabular}{l} $\dot{p_1} (\delta \hat{p_2} +\hat{\delta} \hat{p_0})$ \\ \\ \end{tabular}
\\

\begin{tabular}{l} (0,1,0,1) \\ \\ \end{tabular} & 
\begin{tabular}{l} $\delta (\delta \hat{p_2} +\hat{\delta} \hat{p_0}) (\delta p_3 +\hat{\delta} p_4)$ \\ \\ \end{tabular} &
\begin{tabular}{l} $\delta (\delta \hat{p_2} +\hat{\delta} \hat{p_0}) (\delta \hat{p_1} +\hat{\delta} \hat{p_4})$ \\ \\ \end{tabular} &
\begin{tabular}{l} $\delta^2 p_3 (\delta \hat{p_1} +\hat{\delta} \hat{p_4})$ \\ $+p_0 (\hat{\delta^2} \dot{p_1} +\delta^2 \hat{\delta} p_4)$ \\ \end{tabular} &
\begin{tabular}{l} $\delta^2 \hat{p_3} (\delta \hat{p_1} +\hat{\delta} \hat{p_2})$ \\ $+\hat{p_0} (\hat{\delta^2} \dot{p_1} +\delta^2 \hat{\delta} p_2)$ \\ \end{tabular}
\\ \\

\begin{tabular}{l} (0,1,1,0) \\ \\ \end{tabular} & 
\begin{tabular}{l} $\delta (\delta p_4 +\hat{\delta} p_0) (\delta p_3 +\hat{\delta} p_2)$ \\ \\ \end{tabular} &
\begin{tabular}{l} $\delta (\delta p_4 +\hat{\delta} p_0) (\delta \hat{p_1} +\hat{\delta} \hat{p_2})$ \\ \\ \end{tabular} &
\begin{tabular}{l} $\hat{\delta} (\delta p_4 +\hat{\delta} p_0) (\delta p_3+\dot{p_1})$ \\ \\ \end{tabular} &
\begin{tabular}{l} $\hat{\delta} (\delta \hat{p_2} +\hat{\delta} \hat{p_0}) (\delta p_3 +\dot{p_1})$ \\ \\ \end{tabular}
\\

\begin{tabular}{l} (0,1,1,1) \\ \\ \end{tabular} & 
\begin{tabular}{l} $\delta^2 p_3$ \\ $+\delta \hat{\delta} (p_0 p_2 +\hat{p_0} p_4)$ \\ \end{tabular} &
\begin{tabular}{l} $\delta^2 \hat{p_1}$ \\ $+\delta \hat{\delta} (\hat{p_0} \hat{p_4} +p_0 \hat{p_2})$ \\ \end{tabular} &
\begin{tabular}{l} $\hat{\delta} p_0 (\dot{p_1} +\delta p_3)$ \\ \\ \end{tabular} &
\begin{tabular}{l} $\hat{\delta} \hat{p_0} (\dot{p_1} +\delta p_3)$ \\ \\ \end{tabular}
\\ \\

\begin{tabular}{l} (1,0,0,0) \\ \\ \end{tabular} & 
\begin{tabular}{l} $\hat{\delta} p_0 (\dot{p_2} +\delta p_4)$ \\ \\ \end{tabular} &
\begin{tabular}{l} $\hat{\delta} \hat{p_0} (\dot{p_2} +\delta p_4)$ \\ \\ \end{tabular} &
\begin{tabular}{l} $\delta^2 p_4$ \\ $+\delta \hat{\delta} (\hat{p_0} p_3 +p_0 p_1)$ \\ \end{tabular} &
\begin{tabular}{l} $\delta^2 \hat{p_2}$ \\ $+\delta \hat{\delta} (\hat{p_0} \hat{p_3} +p_0 \hat{p_1})$ \\ \end{tabular}
\\ \\

\begin{tabular}{l} (1,0,0,1) \\ \\ \end{tabular} & 
\begin{tabular}{l} $\delta^2 p_4 (\delta \hat{p_2} +\hat{\delta} \hat{p_3})$ \\ $+p_0 (\hat{\delta^2} \dot{p_2} +\delta^2 \hat{\delta} p_3)$ \\ \end{tabular} &
\begin{tabular}{l} $\delta^2 \hat{p_4} (\delta \hat{p_2} +\hat{\delta} \hat{p_1})$ \\ $+\hat{p_0} (\hat{\delta^2} \dot{p_2} +\delta^2 \hat{\delta} p_1)$ \\ \end{tabular} &
\begin{tabular}{l} $\delta^3 (p_1 p_4 +p_3 \hat{p_4})$ \\ $+\delta \hat{\delta^2} (p_0 p_1 +\hat{p_0} p_3)$ \\ \end{tabular} &
\begin{tabular}{l} $\delta^3 (\hat{p_1} p_2 +\hat{p_2} \hat{p_3})$ \\ $+\delta \hat{\delta^2} (p_0 \hat{p_1} +\hat{p_0} \hat{p_3})$ \\ \end{tabular}
\\ \\

\begin{tabular}{l} (1,0,1,0) \\ \\ \end{tabular} & 
\begin{tabular}{l} $\delta^2 p_2 (\delta p_4 +\hat{\delta} p_3)$ \\ $+p_0 (\hat{\delta} \ddot{p_3} +\delta \hat{\delta^2} p_4)$ \\ \end{tabular} &
\begin{tabular}{l} $\delta^2 \hat{p_2} (\delta p_4 +\hat{\delta} p_1)$ \\ $+\hat{p_0} (\hat{\delta} \ddot{p_1} +\delta \hat{\delta^2} p_4)$ \\ \end{tabular} &
\begin{tabular}{l} $\delta (\delta p_4 +\hat{\delta} p_1) (\delta p_3 +\hat{\delta} p_0)$ \\ \\ \end{tabular} &
\begin{tabular}{l} $\delta (\delta \hat{p_2} +\hat{\delta} \hat{p_1}) (\delta p_3 +\hat{\delta} p_0)$ \\ \\ \end{tabular}
\\ \\

\begin{tabular}{l} (1,0,1,1) \\ \\ \end{tabular} & 
\begin{tabular}{l} $\delta^2 (\hat{p_3} p_4 +p_2 p_3)$ \\ $+\hat{\delta^2} p_0$ \\ \end{tabular} &
\begin{tabular}{l} $\delta^2 (\hat{p_1} \hat{p_4} +p_1 \hat{p_2})$ \\ $+\hat{\delta^2} \hat{p_0}$ \\ \end{tabular} &
\begin{tabular}{l} $\delta^2 p_1 p_3 +\delta \hat{\delta} p_0 p_1$ \\ \\ \end{tabular} &
\begin{tabular}{l} $\delta^2 \hat{p_1} p_3 +\delta \hat{\delta} p_0 \hat{p_1}$ \\ \\ \end{tabular}
\\ \\

\begin{tabular}{l} (1,1,0,0) \\ \\ \end{tabular} & 
\begin{tabular}{l} $\hat{\delta} (\delta p_3 +\hat{\delta} p_0) (\delta p_4 +\dot{p_2})$ \\ \\ \end{tabular} &
\begin{tabular}{l} $\hat{\delta} (\delta \hat{p_1} +\hat{\delta} \hat{p_0}) (\delta p_4 +\dot{p_2})$ \\ \\ \end{tabular} &
\begin{tabular}{l} $\delta (\delta \hat{p_1} +\hat{\delta} \hat{p_0}) (\delta p_4 +\hat{\delta} p_3)$ \\ \\ \end{tabular} &
\begin{tabular}{l} $\delta (\delta \hat{p_1} +\hat{\delta} \hat{p_0}) (\delta \hat{p_2} +\hat{\delta} \hat{p_3})$ \\ \\ \end{tabular}
\\

\begin{tabular}{l} (1,1,0,1) \\ \\ \end{tabular} & 
\begin{tabular}{l} $\dot{p_2} (\delta p_3 +\hat{\delta} p_0)$ \\ \\ \end{tabular} &
\begin{tabular}{l} $\dot{p_2} (\delta \hat{p_1} +\hat{\delta} \hat{p_0})$ \\ \\ \end{tabular} &
\begin{tabular}{l} $\delta p_3 (\delta \hat{p_1} +\hat{\delta} \hat{p_0})$ \\ \\ \end{tabular} &
\begin{tabular}{l} $\delta \hat{p_3} (\delta \hat{p_1} +\hat{\delta} \hat{p_0})$ \\ \\ \end{tabular}
\\

\begin{tabular}{l} (1,1,1,0) \\ \\ \end{tabular} & 
\begin{tabular}{l} $(\delta p_3 +\hat{\delta} p_0) (\delta p_4 +\hat{\delta})$ \\ \\ \end{tabular} &
\begin{tabular}{l} $(\delta \hat{p_1} +\hat{\delta} \hat{p_0}) (\delta p_4 +\hat{\delta})$ \\ \\ \end{tabular} &
\begin{tabular}{l} 0         \\ \\ \end{tabular} &
\begin{tabular}{l} 0         \\ \\ \end{tabular}
\\

\begin{tabular}{l} (1,1,1,1) \\ \\ \end{tabular} & 
\begin{tabular}{l} $\delta p_3 +\hat{\delta} p_0$ \\ \\ \end{tabular} &
\begin{tabular}{l} $\delta \hat{p_1} +\hat{\delta} \hat{p_0}$ \\ \\ \end{tabular} &
\begin{tabular}{l} 0         \\ \\ \end{tabular} &
\begin{tabular}{l} 0         \\ \\ \end{tabular}
\\ \hline
\end{tabular}
} 
\label{tab:M_l}
\end{table}
We then obtain the following result as well as $D(\bm p,\bm q,\bm 1_4)$: 
%
\begin{align}
 (-1)^{\ell} \left<M_{\ell}\right>\ge0\ \ (\ell\in\{1,2,3,4\}). 
\label{eq:detMl}
\end{align}

Determinant $\mathfrak{d}_{\ell}$ is independent of $q_0,q_{\ell}$ but linear to the others. 
Thus, we examined all corner cases of $\bm q_{-0,\ell}\equiv\{q_j;j\in\{1,2,3,4\},j\ne\ell\}$ to check the sign of the expressions listed in Table~\ref{tab:extracted_d}.  
%
\begin{table}[btp]
\centering
\caption{Values of the determinants $\mathfrak{d}_{\ell}$ at the corners of $[0,1]^3$ for $\bm q_{-0,\ell}$; $\theta\equiv T+S,\ \hat{x}\equiv1-x$, $\dot{p_j} \equiv 1 -\delta p_j$, $\ddot{p_j} \equiv 1 -\delta^2 p_j$. }
\scalebox{0.8}[0.8]{
\begin{tabular}{lllll}
\hline
$\bm q_{-0,\ell}$   &   $-\mathfrak{d}_1$   &   $\mathfrak{d}_2$   &   $-\mathfrak{d}_3$   &   $\mathfrak{d}_4$  \\ \hline

\begin{tabular}{l} (0,0,0) \\ \\ \end{tabular} & 
\begin{tabular}{l} $p_1 +\theta \hat{p_1}$ \\ \\ \end{tabular} &
\begin{tabular}{l} $p_3 +\theta \hat{p_3}$ \\ \\ \end{tabular} &
\begin{tabular}{l} $p_2 +\theta \hat{p_2}$ \\ \\ \end{tabular} &
\begin{tabular}{l} $p_4 +\theta \hat{p_4}$ \\ \\ \end{tabular}
\\

\begin{tabular}{l} (0,0,1) \\ \\ \end{tabular} & 
\begin{tabular}{l} $(2 -\theta) p_1 +\delta \hat{p_3}$ \\ $+\delta (p_1 -p_2) +\hat{\delta} \hat{p_1}$ \\ \end{tabular} &
\begin{tabular}{l} $p_3 +\theta \hat{p_3}$ \\ $+\delta (2 -\theta) (p_3 -p_4)$ \\ \end{tabular} &
\begin{tabular}{l} $(2 -\theta) p_2 +\delta \hat{p_3}$ \\ $+\delta \theta (p_1 -p_2) +\hat{\delta} \hat{p_2}$ \\ \end{tabular} &
\begin{tabular}{l} $\delta p_2 +\theta \hat{p_4} +\hat{\delta} p_4$ \\ \\ \end{tabular}
\\ \\

\begin{tabular}{l} (0,1,0) \\ \\ \end{tabular} & 
\begin{tabular}{l} $\theta \hat{p_1} +\delta p_2 +\hat{\delta} p_1$ \\ \\ \end{tabular} &
\begin{tabular}{l} $\delta p_2 +\theta \hat{p_3} +\hat{\delta} p_3$ \\ \\ \end{tabular} &
\begin{tabular}{l} $\theta \hat{p_2} +\delta p_3 +\hat{\delta} p_2$ \\ \\ \end{tabular} &
\begin{tabular}{l} $\delta p_3 +\theta \hat{p_4} +\hat{\delta} p_4$ \\ \\ \end{tabular}
\\

\begin{tabular}{l} (0,1,1) \\ \\ \end{tabular} & 
\begin{tabular}{l} $(2 -\theta) p_1 +\delta \hat{p_3}$ \\ $+\hat{\delta} \hat{p_1}$\\ \end{tabular} &
\begin{tabular}{l} $\delta p_2 +\theta \hat{p_3} +\hat{\delta} p_3$ \\ $+\delta (2 -\theta) (p_3 -p_4)$ \\ \end{tabular} &
\begin{tabular}{l} $\hat{p_2} +(2 -\theta) p_2$ \\ $+\delta \theta (p_1 -p_2)$ \\ \end{tabular} &
\begin{tabular}{l} $\delta p_2 +\theta \hat{p_3}$ \\ $+(\delta+\theta)(p_3-p_4) +\hat{\delta} p_4$ \\ \end{tabular}
\\ \\

\begin{tabular}{l} (1,0,0) \\ \\ \end{tabular} & 
\begin{tabular}{l} $\theta \hat{p_1} +\delta p_3 +\hat{\delta} p_1$ \\ \\ \end{tabular} &
\begin{tabular}{l} $\delta \theta \hat{p_1} +p_3 +\hat{\delta} \theta \hat{p_3}$ \\ \\ \end{tabular} &
\begin{tabular}{l} $\delta \theta \hat{p_1} +p_2 +\hat{\delta} \theta \hat{p_2}$ \\ \\ \end{tabular} &
\begin{tabular}{l} $\delta \theta \hat{p_1} +p_4 +\hat{\delta} \theta \hat{p_4}$ \\ \\ \end{tabular}
\\

\begin{tabular}{l} (1,0,1) \\ \\ \end{tabular} & 
\begin{tabular}{l} $(2 -\theta) p_1 +\delta \hat{p_2}$ \\ $+\hat{\delta} \hat{p_1}$ \\ \end{tabular} &
\begin{tabular}{l} $\delta \hat{p_2} +(2 -\theta) p_3$ \\ $+\hat{\delta} \hat{p_3}$ \\ \end{tabular} &
\begin{tabular}{l} $(2 -\theta) p_2 +\delta \hat{p_3}$ \\ $+\hat{\delta} \hat{p_2}$ \\ \end{tabular} &
\begin{tabular}{l} $\delta \theta \hat{p_1} +\delta p_2$ \\ $+\hat{\delta} \theta \hat{p_4} +\hat{\delta} p_4$ \\ \end{tabular}
\\ \\

\begin{tabular}{l} (1,1,0) \\ \\ \end{tabular} & 
\begin{tabular}{l} $\hat{p_1} +\delta (2 -\theta) p_4$ \\ $+\hat{\delta} (2 -\theta) p_1$ \\ \end{tabular} &
\begin{tabular}{l} $\hat{p_3} +\delta (2 -\theta) p_4$ \\ $+\hat{\delta} (2 -\theta) p_3$ \\ \end{tabular} &
\begin{tabular}{l} $\hat{p_2} +\delta (2 -\theta) p_4$ \\ $+\hat{\delta} (2 -\theta) p_2$ \\ \end{tabular} &
\begin{tabular}{l} $\delta \hat{p_2} +(2 -\theta) p_4$ \\ $+\hat{\delta} \hat{p_4}$\\ \end{tabular}
\\ \\

\begin{tabular}{l} (1,1,1) \\ \\ \end{tabular} & 
\begin{tabular}{l} $\hat{p_1} +(2 -\theta) p_1$ \\ \\ \end{tabular} &
\begin{tabular}{l} $\hat{p_3} +(2 -\theta) p_3$ \\ \\ \end{tabular} &
\begin{tabular}{l} $\hat{p_2} +(2 -\theta) p_2$ \\ \\ \end{tabular} &
\begin{tabular}{l} $\hat{p_4} +(2 -\theta) p_4$ \\ \\ \end{tabular}
\\ \hline
\end{tabular}
} 
\label{tab:extracted_d}
\end{table}
Note that $\theta\equiv T+S$, which satisfies $0<\theta<2$, owing to Eq.~(\ref{eq:PD_cond}). 
For Table~\ref{tab:extracted_d}, we also obtain the following result: 
\begin{align}
 (-1)^{\ell} \mathfrak{d}_{\ell}>0\ \ (\ell\in\{1,2,3,4\}). 
\label{eq:frakD}
\end{align}

Therefore, we obtain the following lemma through Eqs.~(\ref{eq:detMl}) and (\ref{eq:frakD}): 

%
\begin{lemma}
\begin{oframed}
For pcZD strategy $\bm p$ with discount factor $\delta\ (\delta_c<\delta<1)$, 
%
\begin{align}
 \frac{\partial s_Y}{\partial q_1}, \frac{\partial s_Y}{\partial q_2}, \frac{\partial s_Y}{\partial q_3}, \frac{\partial s_Y}{\partial q_4} \ge 0, 
\end{align}
for every $\bm q$ (all conditions such that $\partial s_Y/\partial q_{\ell}=0$ are written in Appendix~\ref{app:partial1234_conds}). 
\label{lem:partial1234}
\end{oframed}
\end{lemma}

By contrast, Table~\ref{tab:q0} shows the result of $\mathfrak{d_0}$, which is independent of $q_0$ but linear to every $q_{\ell}$. 
%
\begin{table}[btp]
\centering
\caption{Values of the determinants $\mathfrak{d}_0$ at the corners of $[0,1]^4$ for $\{q_1,q_2,q_3,q_4\}$; $\hat{x}\equiv1-x$, $\dot{p_j} \equiv 1 -\delta p_j$, $\ddot{p_j} \equiv 1 -\delta^2 p_j$. }
\scalebox{0.85}[0.85]{
\begin{tabular}{lllll}
\hline
$(q_1,q_2)$ & $(q_3,q_4)$ &       &       &       \\ \cline{2-5} 
            & (0,0)       & (0,1) & (1,0) & (1,1) \\ \hline
%
\\
\begin{tabular}{l} (0,0) \\ \\ \end{tabular} &
\begin{tabular}{l} $p_0 +\theta \hat{p_0}$ \\ \\ \end{tabular} & 
\begin{tabular}{l} $\delta (\theta -1) p_1 +\delta (p_1 -p_2) +\dot{p_3}$ \\ $+(1 -\theta +\delta (2 -\theta)) p_0$ \\ \end{tabular}   &
\begin{tabular}{l} $\delta p_2 +\theta \hat{p_0} +\hat{\delta} p_0$ \\ \\ \end{tabular}   &
\begin{tabular}{l} $\delta \theta p_1 +\dot{p_3}$ \\ $+(1 +\delta) (1 -\theta) p_0$ \\ \end{tabular}
\\ \\
\begin{tabular}{l} (0,1) \\ \\ \end{tabular} &
\begin{tabular}{l} $\delta p_3 +\theta \hat{p_0}$ \\ $+\hat{\delta} p_0$ \\ \end{tabular}   &
\begin{tabular}{l} $\delta \theta p_1 +\dot{p_2} $ \\ $+(1 +\delta) (1 -\theta) p_0$ \\ \end{tabular}   &
\begin{tabular}{l} $\theta +\delta p_2 +\delta p_3$ \\ $+(1 -2\delta -\theta) p_0$ \\ \end{tabular}   &
\begin{tabular}{l} $1 +\delta \theta p_1$ \\ $+(1 -(1 +\delta) \theta) p_0$ \\ \end{tabular}
\\ \\
\begin{tabular}{l} (1,0) \\ \\ \end{tabular} &
\begin{tabular}{l} $\theta (\hat{\delta} \hat{p_0} +\delta \hat{p_1})$ \\ $+p_0$ \\ \end{tabular} &
\begin{tabular}{l} $(2-\theta) p_0 +\delta (\hat{p_2} +\hat{p_3})$ \\ $+(1-2\delta) \hat{p_0}$ \\ \end{tabular}   &
\begin{tabular}{l} $(2-\theta) (\hat{\delta} p_0 +\delta p_4)$ \\ $+\hat{\delta} \hat{p_0} +\delta \hat{p_3}$ \\ \end{tabular}   &
\begin{tabular}{l} $(2-\theta) p_0 $ \\ $+\hat{\delta} \hat{p_0} +\delta \hat{p_3}$ \\ \end{tabular}
\\ \\
\begin{tabular}{l} (1,1) \\ \\ \end{tabular} &
\begin{tabular}{l} $\theta (\hat{\delta} \hat{p_0} +\delta \hat{p_1})$ \\ $+\hat{\delta} p_0 +\delta p_3$ \\ \end{tabular}   &
\begin{tabular}{l} $(2-\theta) p_0 +\hat{\delta} \hat{p_0} +\delta \hat{p_2}$ \\ \\ \end{tabular}   &
\begin{tabular}{l} $(2 -\theta) (\hat{\delta} p_0 +\delta p_4) +\hat{p_0}$ \\ \\ \end{tabular} &
\begin{tabular}{l} $(2 -\theta) p_0 +\hat{p_0}$ \\ \\ \end{tabular}
\\ \hline
\end{tabular}
} 
\label{tab:q0}
\end{table}
Analyzing Table~\ref{tab:q0}, we found that in some corner cases (e.g., $(q_1,q_2,q_3,q_4)=(0,1,1,0)$) $\mathfrak{d}_0$ can take a negative value. 
Thus, $\partial s_Y/\partial q_0$ also occasionally becomes negative, and thus it is difficult to predict the change in $q_0$. 


However, it is clear that each $q_{\ell}$ increases regardless of $q_0$ because of Lemma~\ref{lem:partial1234}. 
\highlight{
Such increases of $q_{\ell}$ continue until $\partial s_Y/\partial q_{\ell}$ reaches 0 or $q_{\ell}=1$. 
That is, if $\bm q'=(q'_0; q'_1, q'_2, q'_3, q'_4)$ is the state at which all $q_{\ell}$ increases have stopped, it satisfies the following condition: 
\begin{align}
 \left. \frac{\partial s_Y}{\partial q_{\ell}} \right|_{\bm q'} = 0\ \ \forall \ell \in \{1,2,3,4\}\ \ s.t.\ \ q'_{\ell}<1. 
 \label{eq:cond1}
\end{align}
}
Because $q_0$ may be able to improve, $\bm q'$ may not be the final strategy of an adapting path. 
Thus, we refer to this as relay points. 

We explored $\bm q'$ by examining all corner cases. 
Table~\ref{tab:all_corner} shows the condition under which the value of $\partial s_X/\partial q_{\ell}$ at each corner case is 0. 
Note that the checked corner cases in Table~\ref{tab:all_corner} can satisfy Eq.~(\ref{eq:cond1}) (occasionally under the condition of $\bm p$). 
%
\begin{landscape}
\begin{table}[hbtp]
\newcommand{\expandrow}{\rule[-7mm]{0mm}{10mm}}

\centering
\caption{
Conditions under which the value of $\partial s_X/\partial q_{\ell}$ at each corner case is 0. 
``None" indicates that the partial derivative is not 0 regardless of $q_j\ (j\in\{0,1,2,3,4\})$, 
``All" means that the partial derivative is always 0, and ``*" indicates that $q_{\ell}$ has already become 1.  
Moreover, the checked corner cases means it satisfies Eq.~(\ref{eq:cond1}) (occasionally under a condition of $\bm p$). 
}
\scalebox{0.75}[0.75]{
\begin{tabular}{rlllllp{10mm}rlllll}
\hline
Eq.~(\ref{eq:cond1})  &  $\bm q$   &   $\displaystyle \frac{\partial s_X}{\partial q_1}$    &   $\displaystyle \frac{\partial s_X}{\partial q_2}$   &   $\displaystyle \frac{\partial s_X}{\partial q_3}$   &   $\displaystyle \frac{\partial s_X}{\partial q_4}$    &      &
Eq.~(\ref{eq:cond1})  &  $\bm q$   &   $\displaystyle \frac{\partial s_X}{\partial q_1}$    &   $\displaystyle \frac{\partial s_X}{\partial q_2}$   &   $\displaystyle \frac{\partial s_X}{\partial q_3}$   &   $\displaystyle \frac{\partial s_X}{\partial q_4}$    \\  \hline
  &  00000   & All  & All  & $(p_0,p_4)=(0,0)$  & None  &  &\expandrow
  &  10000   & $p_0=0$  & $p_0=1$  & None  & None  \\
  &  00001   & $p_4=0$  & None  & $(p_0,p_4)=(0,0)$  & *  &  &\expandrow
  &  10001   & $(p_0,p_4)=(0,0)$  & None  & None  & *  \\
  &  00010   & \begin{tabular}[c]{@{}l@{}}$p_2=0$ or\\ $(p_0,p_4)=(0,0)$\end{tabular}   & $(p_0,p_4)=(0,0)$  & *  & None  &   &\expandrow
  &  10010   & $(p_0,p_2)=(0,0)$  & None  & *  & $(p_1,p_3)=(1,1)$  \\
  &  00011   & \begin{tabular}[c]{@{}l@{}}$(p_2,p_4)=(0,0)$ or\\ $(p_0,p_2,p_3)=(1,0,1)$\end{tabular} & None  & *  & *  &  &\expandrow
  &  10011   & \begin{tabular}[c]{@{}l@{}}$(p_0,p_2,p_3)=(0,0,1)$ or\\ $(p_0,p_2,p_4)=(0,0,0)$\end{tabular} & None  & *  & *  \\
  &  00100   & All  & *  & $(p_0,p_4)=(0,0)$  & None  &  &\expandrow
  &  10100   & None  & *  & None  & None  \\
  &  00101   & None  & *  & None  & *  &  &\expandrow
  &  10101   & None  & *  & None  & *  \\
  &  00110   & $(p_0,p_4)=(0,0)$  & *  & *  & \begin{tabular}[c]{@{}l@{}}$(p_0,p_1,p_3)=(1,1,1)$ or\\ $(p_0,p_2,p_3)=(1,0,1)$\end{tabular} &  &\expandrow
  &  10110   & None  & *  & *  & $p_1=1$  \\
  &  00111   & None  & *  & *  & * &  &\expandrow
  &  10111   & None   & *  & *  & *  \\
  &  01000   & * & All   &   $(p_0,p_4)=(0,0)$  & None  &  &\expandrow
$\checkmark$  &  11000   & *  & $(p_0,p_1)=(1,1)$  & $(p_0,p_1)=(1,1)$  & $(p_0,p_1)=(1,1)$  \\
  &  01001   & *  & None  & None  & *  &  &\expandrow
$\checkmark$  &  11001   & *  & $(p_0,p_1)=(1,1)$  & $(p_0,p_1)=(1,1)$  & *  \\
  &  01010   & *  & $(p_0,p_4)=(0,0)$  & *  & $(p_0,p_3)=(1,1)$  &  &\expandrow
$\checkmark$  &  11010   & *  & $(p_0,p_1)=(1,1)$  & *  & \begin{tabular}[c]{@{}l@{}}$(p_0,p_1)=(1,1)$ or\\ $p_3=1$\end{tabular} \\
  &  01011   & * & None  & *  & *  &  &\expandrow
$\checkmark$  &  11011   & * & $(p_0,p_1)=(1,1)$  & *  & * \\
  &  01100   & *  & *  & $(p_0,p_4)=(0,0)$  & None  &  &\expandrow
$\checkmark$  &  11100   & *  & *  & All  & All  \\
$\checkmark$  &  01101   & *  & *  & $p_0=0$  & *  &  &\expandrow
$\checkmark$  &  11101   & *  & *  & All  & *  \\
$\checkmark$  &  01110   & *  & *  & *  & $p_0=1$  &  &\expandrow
$\checkmark$  &  11110   & *  & *  & *  & All  \\
$\checkmark$  &  01111   & *  & *  & *  & *  &  &\expandrow
$\checkmark$  &  11111   & * & *  & *  & *  \\
%
%
\hline
\end{tabular}
}
\label{tab:all_corner}
\end{table}
\end{landscape}

Summarizing all cases that satisfy Eq.~(\ref{eq:cond1}), we obtain the following lemma: 
%
\begin{oframed}
\begin{lemma}
For pcZD strategy $\bm p$ with discount factor $\delta\ (\delta_c<\delta<1)$, every $\bm q'$ satisfies one of the following conditions: 
\begin{enumerate}
\renewcommand{\labelenumi}{{\rm (R\arabic{enumi})}}
\item $(q'_0, q'_1, q'_2) = (1,1,1)$				\label{enum:sub1}
\item $(p_0, p_1, q'_0, q'_1) = (1,1,1,1)$			\label{enum:sub2}
\item $(p_0, q'_1, q'_2, q'_3, q'_4) = (0,1,1,0,1)$	\label{enum:sub3}
\item $(p_0, q'_1, q'_2, q'_3, q'_4) = (1,1,1,1,0)$	\label{enum:sub4}
\item $(q'_1, q'_2, q'_3, q'_4) = (1,1,1,1)$		\label{enum:sub5}
\end{enumerate}
\label{lem:substationary}
\end{lemma}
\end{oframed}


Because player $Y$ reaches $\bm q'$ regardless of $q_0$, it is possible to improve $q_0$. 
We examined the sign of $\partial s_X/\partial q_0$ again for each condition of Lemma~\ref{lem:substationary}: 

\begin{itemize}
\item For the condition (R\ref{enum:sub1}), we obtain $\mathfrak{d}_0>0$ because every expression of the bottom row in Table~\ref{tab:q0} is greater than 0, that is, $\partial s_Y/\partial q_0>0$; 
\item For the condition (R\ref{enum:sub2}), we obtain $\mathfrak{d}_0>0$ by substituting $(p_0,p_1)=(1,1)$ into the bottom two rows (shown in Table~\ref{tab:partial0_S2}), that is, $\partial s_Y/\partial q_0>0$; 
\item For the condition (R\ref{enum:sub3}), we obtain $\mathfrak{d}_0=\hat{\delta}+\delta\hat{p_2}>0$ from the expression of $(q_1,q_2,q_3,q_4)=(1,1,0,1)$ in Table~\ref{tab:q0}, where $p_0=0$, that is, $\partial s_Y/\partial q_0>0$; 
\item For the conditions (R\ref{enum:sub4}) and (R\ref{enum:sub5}), we obtain $\mathfrak{d}_0>0$ along with (R\ref{enum:sub3}), that is, $\partial s_Y/\partial q_0>0$. 
\end{itemize}
%
\begin{table}[]
\centering
\caption{Values of the determinants $\mathfrak{d}_0$ at the corners of $[0,1]^3$ for $(q_2,q_3,q_4)$ where $(p_0,p_1)=(1,1)$. }
\scalebox{0.85}[0.85]{
\begin{tabular}{lllll}
\hline
$q_2$ & $(q_3,q_4)$ &       &       &       \\ \cline{2-5} 
            & (0,0)       & (0,1) & (1,0) & (1,1) \\ \hline
%
0  &  $(2-\theta) (\hat{\delta} +\delta p_4) +\delta (\hat{p_2} +\hat{p_3})$  &  $2-\theta +\delta \hat{p_2} +\delta \hat{p_3}$  &  $(2-\theta) (\hat{\delta} +\delta p_4) +\delta \hat{p_3}$  &  $2-\theta +\delta \hat{p_3}$  \\
1  &  $(2-\theta) (\hat{\delta} +\delta p_4) +\delta \hat{p_2}$  &  $2-\theta +\delta \hat{p_2}$  &  $(2-\theta) (\hat{\delta} +\delta p_4)$  &  $2-\theta$  \\
\hline
\end{tabular}
} 
\label{tab:partial0_S2}
\end{table}

From the above discussion, we found that $\partial s_Y/\partial q_0>0$ for every condition of Lemma~\ref{lem:substationary}; thus, player $Y$ can improve $s_Y$ by increasing $q_0$ unless $q_0=1$ after going through the relay points. 
This is sustained unless $\bm q$ satisfies the following condition along with $q_{\ell}$, and finally reaches $\bm q^*=(q^*_0,q^*_1,q^*_2,q^*_3,q^*_4)$: 
\begin{align}
\left. \frac{\partial s_Y}{\partial q_j} \right|_{\bm q^*} = 0 \ \ \forall j\in\{0,1,2,3,4\} \mathrm{\ \ s.t.\ \ } q^*_j<1. 
\end{align}

Referring to Lemma~\ref{lem:substationary}, player $Y$ cannot increase $q_0$ because $q_0=1$ is already satisfied for (R\ref{enum:sub1}) and (R\ref{enum:sub2}). 
By contrast, for (R\ref{enum:sub3}) and (R\ref{enum:sub4}), player $Y$ can increase $q_0$, and finally, $\bm q^*$ satisfies (R\ref{enum:sub1}). 
For (R\ref{enum:sub5}), player $Y$ can increase $q_0$ if $q'_0<1$, and finally $\bm q^*$ reaches (R\ref{enum:sub1}). 

$\therefore$ Because there are no singularities over $\bm q\in[0,1]^5$ as proved by Lemma~\ref{corollary:singular}, all adapting paths of the adaptive player $Y$ lead to the strategies satisfying one of the following conditions for the pcZD strategy of player $X$ with discount factor $\delta\in(0,1)$:  
\begin{enumerate}
\renewcommand{\labelenumi}{(T\arabic{enumi})}
\item $(q^*_0, q^*_1, q^*_2) = (1,1,1)$, 
\item $(p_0, p_1, q^*_0, q^*_1) = (1,1,1,1)$. 
\end{enumerate}

\begin{figure}[btp]
\centering
\includegraphics[width=\columnwidth]{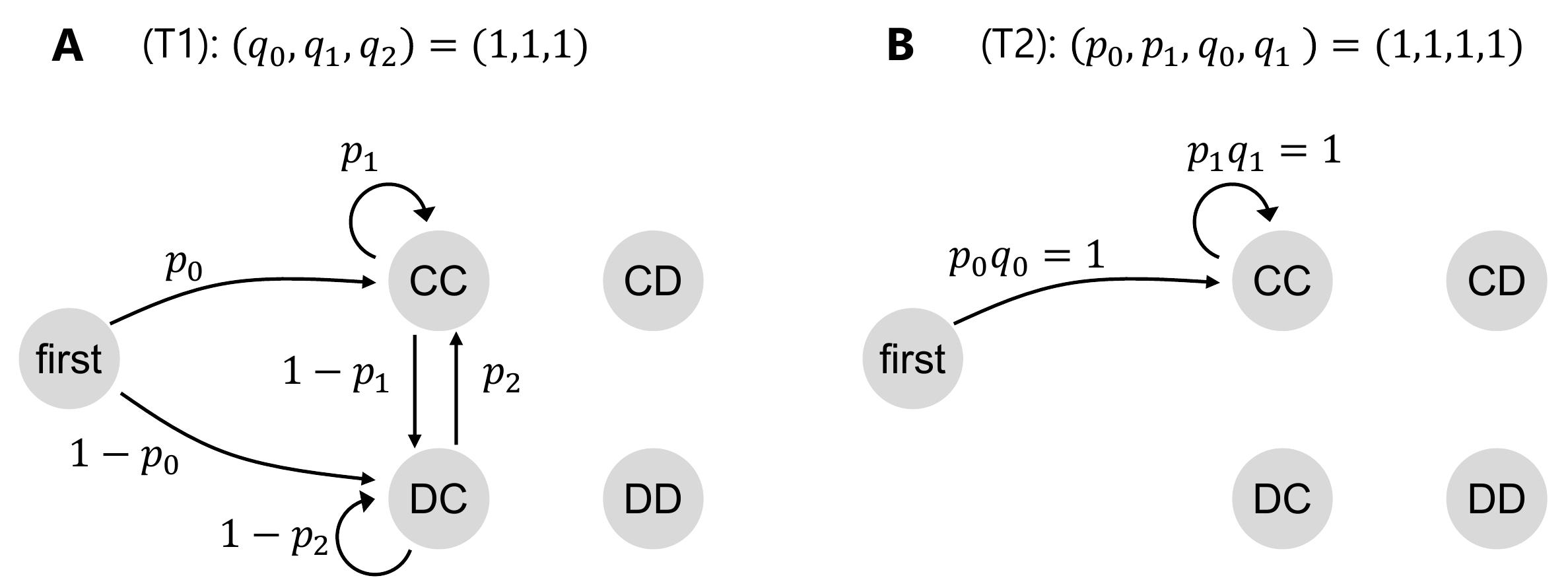}
\caption{
Transition diagram for (T1) and (T2). 
The ``first" state represents the first round before the players act. 
The remaining states represent the actions between players $X$ and $Y$. 
The arrows show possible transitions between states, each of which has a transition probability. 
}
\label{fig:transition}
\end{figure}

We show the transition diagrams for (T1) and (T2) in Fig.~\ref{fig:transition}. 
If $\bm q$ satisfies (T1), then player $Y$ always cooperates in the first round because of $q_0=1$, and continues to cooperate in the subsequent rounds regardless of player $X$'s action owing to $(q_1,q_2)=(1,1)$ (Fig.~\ref{fig:transition}A). 
Moreover, if $\bm p,\bm q$ satisfy (T2), then both players cooperate in the first round because of $(p_0,q_0)=(1,1)$, and continue to cooperate in the subsequent rounds because of $(p_1,q_1)=(1,1)$ (Fig.~\ref{fig:transition}B). 
Therefore, a pcZD strategy leads player $Y$ to an unconditional cooperation. 

\section{Numerical examples}
Herein, we show the numerical simulations how an adaptive player changes its strategy over time. 
%

Adaptive player $Y$ updates its strategy $\bm q$ to satisfy (A1) of Def.~\ref{def:adapting_path} 
each time players finish executing the game a sufficient number of times. 
\highlight{
We define the update rule from $\bm q^n=(q_0^n; q_1^n, q_2^n, q_3^n, q_4^n)$ to $\bm q^{n+1}=(q_0^{n+1}; q_1^{n+1}, q_2^{n+1}, q_3^{n+1}, q_4^{n+1})$ using gradient ascent below: 
\begin{align}
q_{j}^{n+1} = \max\left(\min\left(q_j^n +\nu \left.\frac{\partial s_Y}{\partial q_j}\right|_{\bm q=\bm q^n}, 1\right), 0\right), 
\label{eq:diffeq}
\end{align}
where $\nu > 0$ means the learning rate of player $Y$. 

The partial derivatives in the right side of Eq.~\eqref{eq:diffeq} were obtained through the centered difference approximation. 
The calculation algorithm is shown in Alg.~\ref{alg:ga}. 
Note that we used $\Delta q=0.0001$, and $\nu=0.1$. 
}

\begin{algorithm}
\begin{multicols}{2}
\setstretch{1.5} 
\DontPrintSemicolon

\SetKwFunction{FMain}{$f$}
\SetKwProg{Fn}{Function}{:}{}
\nl \Fn{\FMain{$\bm q$, $j$, $\nu$, $\Delta q$}}{
	\nl define $\bm e_0=(1,0,0,0,0)$\;
	\nl define $\bm e_1=(0,1,0,0,0)$\;
	\nl define $\bm e_2=(0,0,1,0,0)$\;
	\nl define $\bm e_3=(0,0,0,1,0)$\;
	\nl define $\bm e_4=(0,0,0,0,1)$\;
	\;
	\nl $\bm q^{+}=\bm q+\Delta q \bm e_j$\;
	\nl $\bm q^{-}=\bm q-\Delta q \bm e_j$\;
	\nl calculate $s^+_Y$ and $s^-_Y$ using $\bm q^+$ and $\bm q^-$, respectively\;
	\nl \textbf{return} $\nu\displaystyle{\frac{s^+_Y-s^-_Y}{2\Delta q}}$\;
}

\nl $n \leftarrow 0$\;
\nl generate random vector $\bm q_0\in[0,1]^5$\;
\nl $\bm q^n = \bm q_0$\;
\nl set $\bm p$ as a pcZD strategy\;
\nl\While{True} {
	\nl\For{$j\leftarrow0$ \KwTo $4$} {
		\nl $q_j^{n+1} = \min\{\max\{q_j^{n}+f(\bm q^n, j, \nu, \Delta q), 0\}, 1\}$\;
	}
\;
	\tcc{finish this iteration if $\bm q^{n+1}$ is nearly equal to $\bm q^n$}
	\nl\If{$|\bm q^{n+1}-\bm q^n|<10^{-12}$}{\label{ut}
		\nl break\;
	}
	\nl $n\leftarrow n+1$\;
}
\end{multicols}
\caption{Calculation algorithm}
\label{alg:ga}
\end{algorithm}

\begin{figure}[btp]
\centering
\includegraphics[width=\columnwidth]{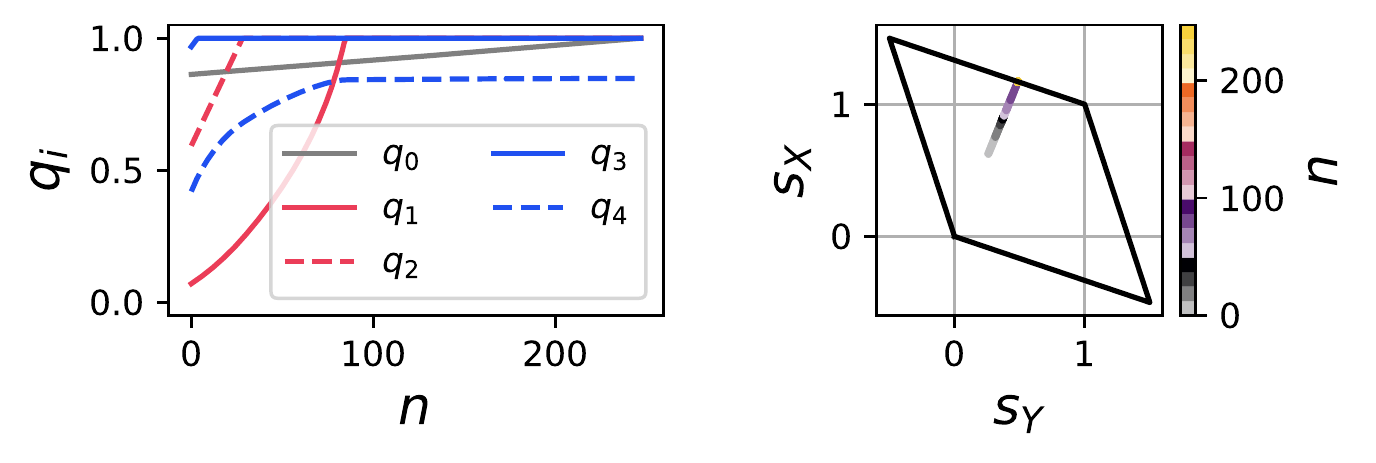}
\caption{
Improvement of $q_j$ (left) and $(s_Y,s_X)$ as time goes by, where $\delta=0.99,\ (T,S)=(1.5,-0.5)$, $\bm q_0=(0.863;0.071,0.593,0.963,0.420)$, and $\bm p=(0.0; 0.75, 0.25, 0.5, 0.0)$. 
In the left figure, the five lines show the changes in $\bm q$: gray line is $q_0$; red solid line is $q_1$; red dashed line is $q_2$; blue solid line is $q_3$; blue dashed line is $q_4$. 
The simulation ended at $n=247$. 
In the right figure, the color of the points indicates the change in time.  
}
\label{fig:each_qi_99}
\end{figure}

\begin{figure}[btp]
\centering
\includegraphics{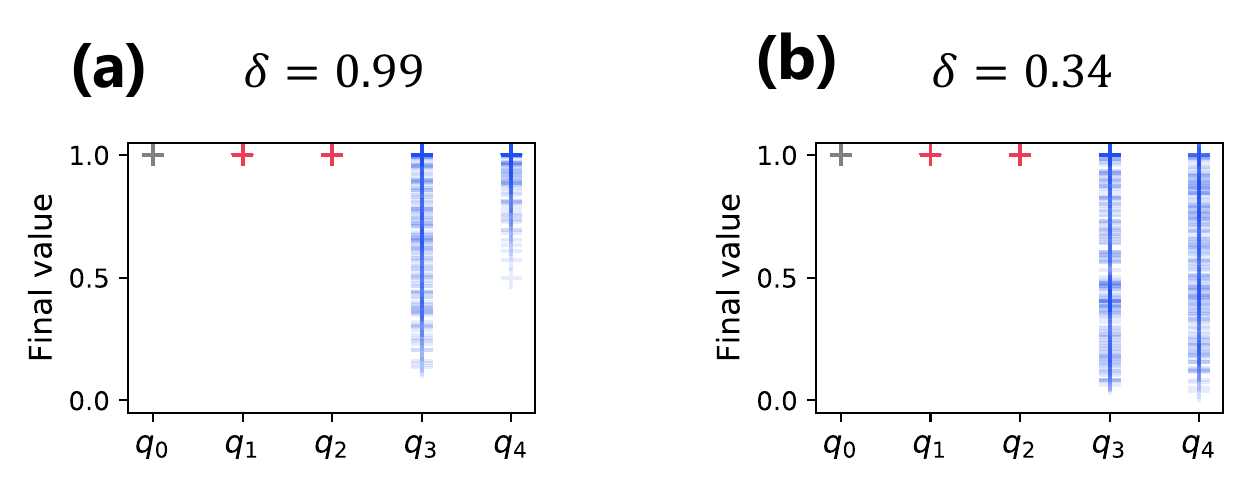}
\caption{
Final strategies obtained from the various initial strategies (which are 100 random instances): 
{\bf (a)} shows the result where $(T,S)=(1.5,-0.5),\ \delta=0.99,\ \bm p=(0.0; 0.75, 0.25, 0.5, 0.0)$;  
{\bf (b)} shows the result where $(T,S)=(1.5,-0.5),\ \delta=0.34,\ \bm p=(0.0; 1.0, 0.0, 1.0, 0.0)$. 
Each point shows the value of $q_j$ at the point where the simulation ends. 
}
\label{fig:finals}
\end{figure}

In Fig.~\ref{fig:each_qi_99}, we show the change in each element in $\bm q$ (left) and both of the expected payoffs $(s_Y,s_X)$ (right), where $\delta=0.99,\ (T,S)=(1.5,-0.5)$, $\bm q_0=(0.863;0.071,0.593,0.968,0.420)$, and player $X$ uses pcZD strategy $\bm p=(0.0; 0.75, 0.25, 0.5, 0.0)$. 
%
\begin{figure}[btp]
\centering
\includegraphics[width=\columnwidth]{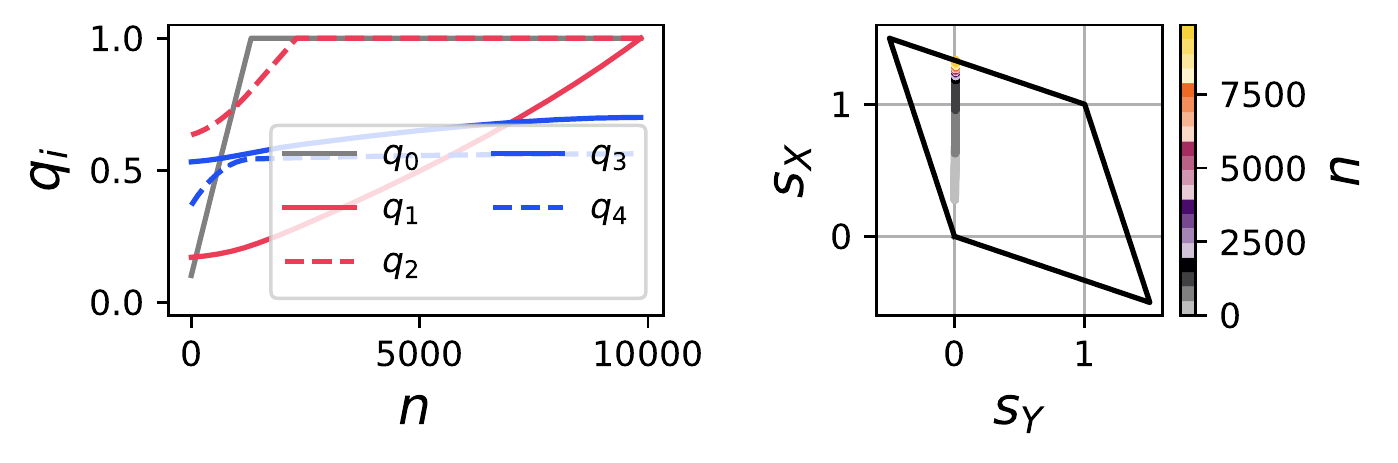}
\caption{
Improvement of $q_j$ (left) and $(s_Y,s_X)$ as time goes by, where $\delta=0.34,\ (T,S)=(1.5,-0.5)$, $\bm q(0)=(0.102;0.171,0.634,0.532,0.368)$, and $\bm p=(0.0; 1.0, 0.0, 1.0, 0.0)$. 
In the left figure, the five lines show the changes in $\bm q$: gray line is $q_0$; red solid line is $q_1$; red dashed line is $q_2$; blue solid line is $q_3$; blue dashed line is $q_4$. 
The simulation ended at $n=9841$. 
In the right figure, the color of the points indicates the change in time.  
}
\label{fig:each_qi_34}
\end{figure}
We can see that all values of $\bm q$ increase, and $q_0$, $q_1$, and $q_2$ reach 1, which denotes that the condition (T1) is satisfied. 
From the right part of Fig.~\ref{fig:each_qi_99}, both $s_Y$ and $s_X$ reach the maximized value.
\highlight{
We also show the result using almost the lowest discount factor $\delta$ at which pcZD can exist, i.e. $\delta=0.34$, in Fig.~\ref{fig:each_qi_34}. 
Note that we have $\delta_c=1/3$ from Eq.~\eqref{eq:delta_c} when $(T,S)=(1.5,-0.5)$. 
In this case, as well, we can find that the adaptive player adapted its strategy until satisfying the condition (T1). 
Furthermore, we obtained Fig.~\ref{fig:finals} which shows the final values of $\bm q$ for each of the 200 random initial strategies in both cases: (a) $\delta=0.99$ and (b) $\delta=0.34$. 
}
%
As both figures indicate,  $q_0$, $q_1$, and $q_2$ always reach 1, and $q_3$ and $q_4$ occasionally stop increasing at the intermediate value. 

\begin{figure}[btp]
\centering
\includegraphics[width=\columnwidth+10mm]{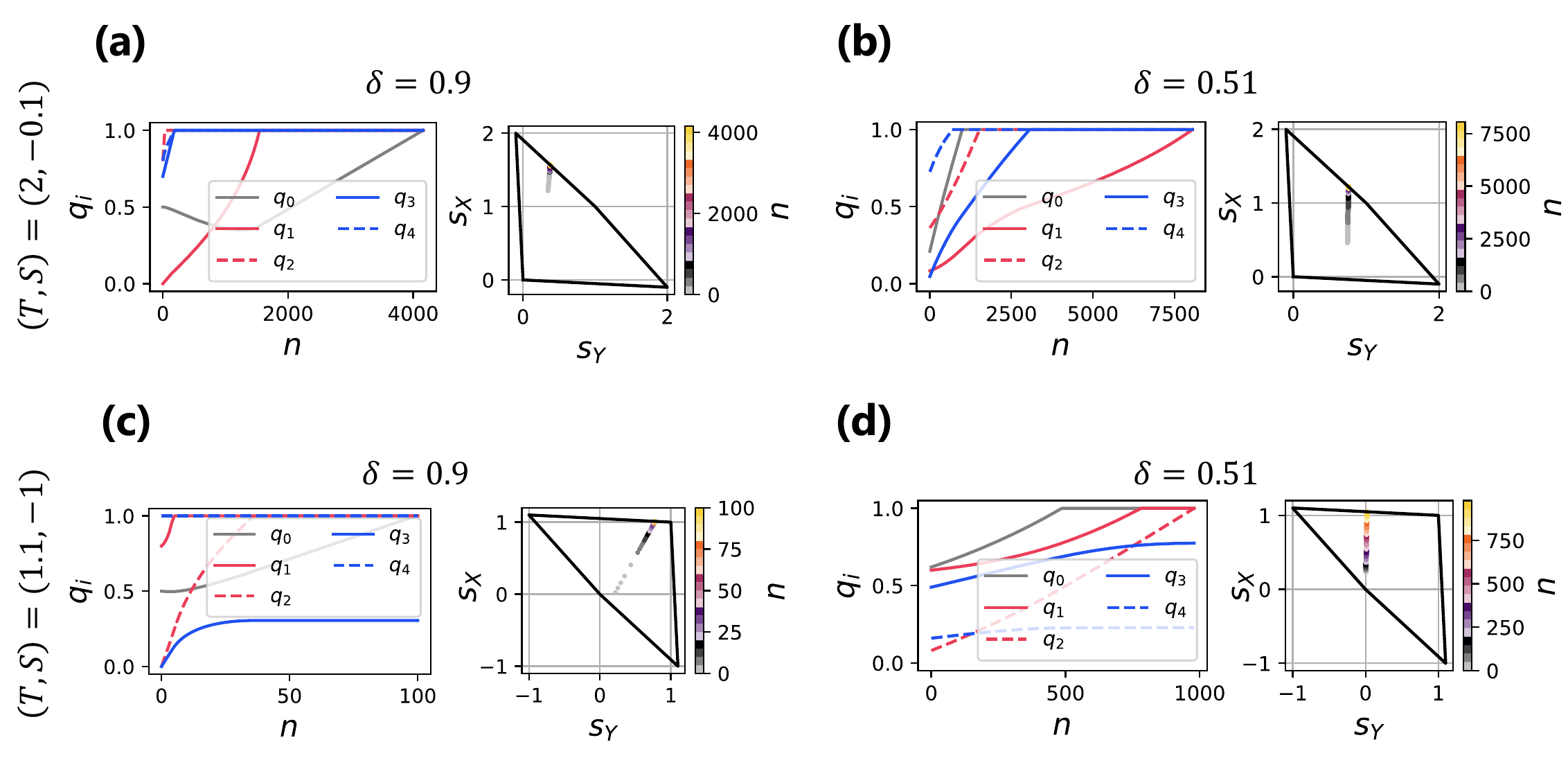}
\caption{
Improvement of $q_j$ (left) and $(s_Y,s_X)$ as time goes by, where 
{\bf (a)} $(T,S)=(2.0,-0.1),\ \delta=0.9$, $\bm q(0)=(0.5;0.0,0.8,0.7,0.8)$, and $\bm p=(0.95;0.7,0.2,0.13,0.0)$; %
{\bf (b)} $(T,S)=(2.0,-0.1),\ \delta=0.51$, $\bm q(0)=(0.211;0.084,0.362,0.047,0.726)$, and $\bm p=(0.750;1.0,0.0,0.135,0.0)$; %
{\bf (c)} $(T,S)=(1.1,-1.0),\ \delta=0.9$, $\bm q(0)=(0.5;0.8,0.0,0.0,1.0)$, and $\bm p=(0.0;1.0,0.4,0.89,0.1)$; %
and {\bf (d)} $(T,S)=(1.1,-1.0),\ \delta=0.51$, $\bm q(0)=(0.62;0.6,0.08,0.49,0.16)$, and $\bm p=(0.0;1.0,0.865,1.0,0.0)$. %
On the left side of each figure, the five lines show the changes in $\bm q$: gray line is $q_0$; red solid line is $q_1$; red dashed line is $q_2$; blue solid line is $q_3$; blue dashed line is $q_4$. 
On the right side, the color of the points indicates the change in time.  
}
\label{fig:T1_each_summary}
\end{figure}

\begin{figure}[btp]
\centering
\includegraphics[width=\columnwidth]{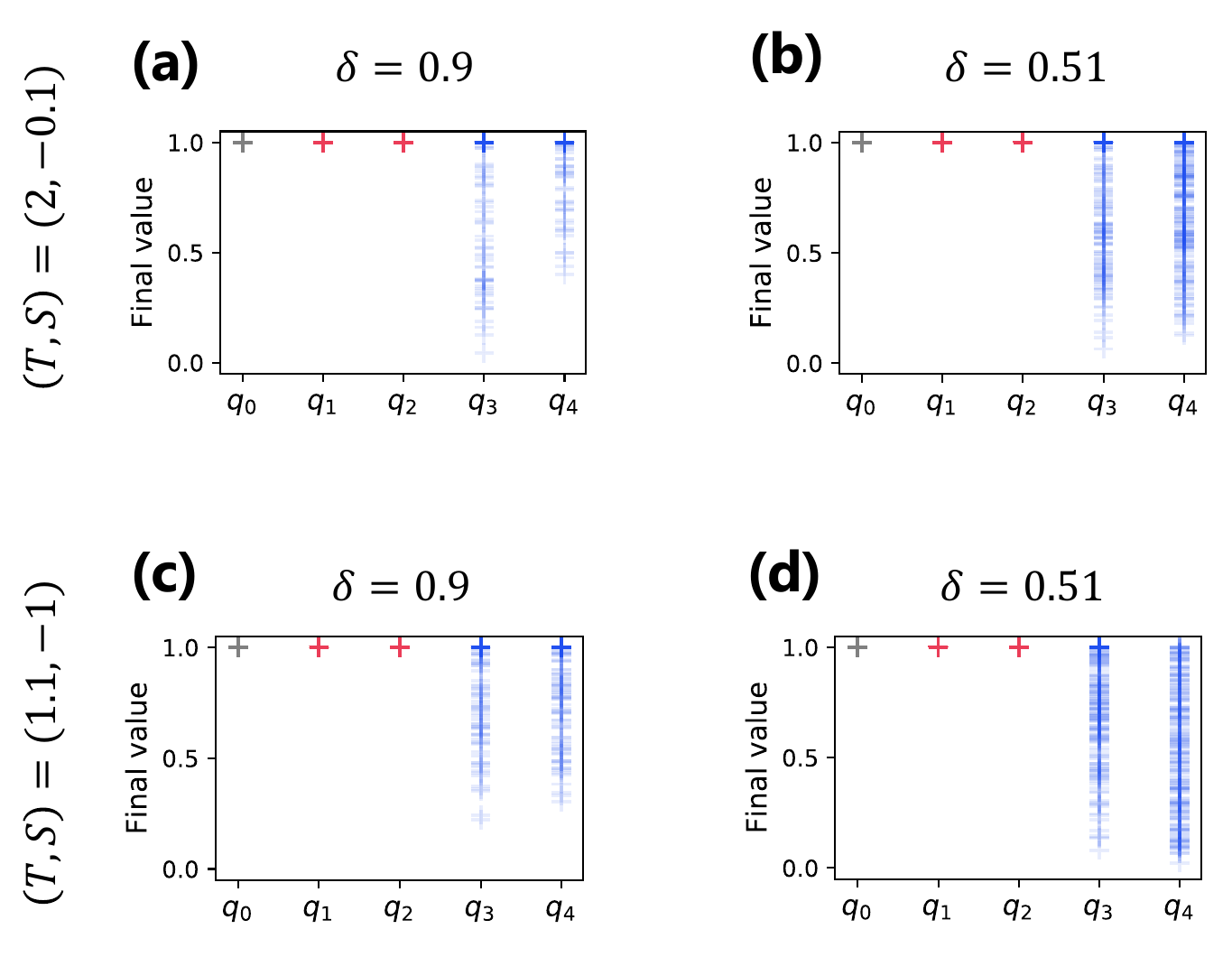}
\caption{
Final strategies obtained from the various initial strategies (which are 100 random instances), where %
\highlight{
{\bf (a)} $(T,S)=(2.0,-0.1),\ \delta=0.99$, and $\bm p=(0.95;0.8,0.2,0.411,0.0)$; %
{\bf (b)} $(T,S)=(2.0,-0.1),\ \delta=0.51$, and $\bm p=(0.750;1.0,0.0,0.135,0.0)$; %
{\bf (c)} $(T,S)=(1.1,-1.0),\ \delta=0.99$, and $\bm p=(0.0;0.7,0.37,0.799,0.1)$; %
and {\bf (d)} $(T,S)=(1.1,-1.0),\ \delta=0.51$, and $\bm p=(0.5;1.0,0.865,1.0,0.0)$. %
}
Each point shows the value of $q_j$ at the point where the simulation ends. %
}
\label{fig:T1_final_summary}
\end{figure}

\highlight{
To reveal the robustness of our result, we simulated with other values of $(T, S)$ and $\delta$. 
Figures~\ref{fig:T1_each_summary} and~\ref{fig:T1_final_summary} provide the results for some cases of the combination of $(T, S)$ and $\delta$. 
As in Figs.~\ref{fig:each_qi_99}, \ref{fig:each_qi_34}, and \ref{fig:finals}, all of these results show that all adapting paths reach the final states satisfying the condition (T1). 
Thereby, we can confirm the robustness for (T1).
Note that we showed the special case in which $q_0$ initially decreased in Fig.~\ref{fig:T1_each_summary}(a). 
}

\begin{figure}[btp]
\centering
\includegraphics[width=\columnwidth+10mm]{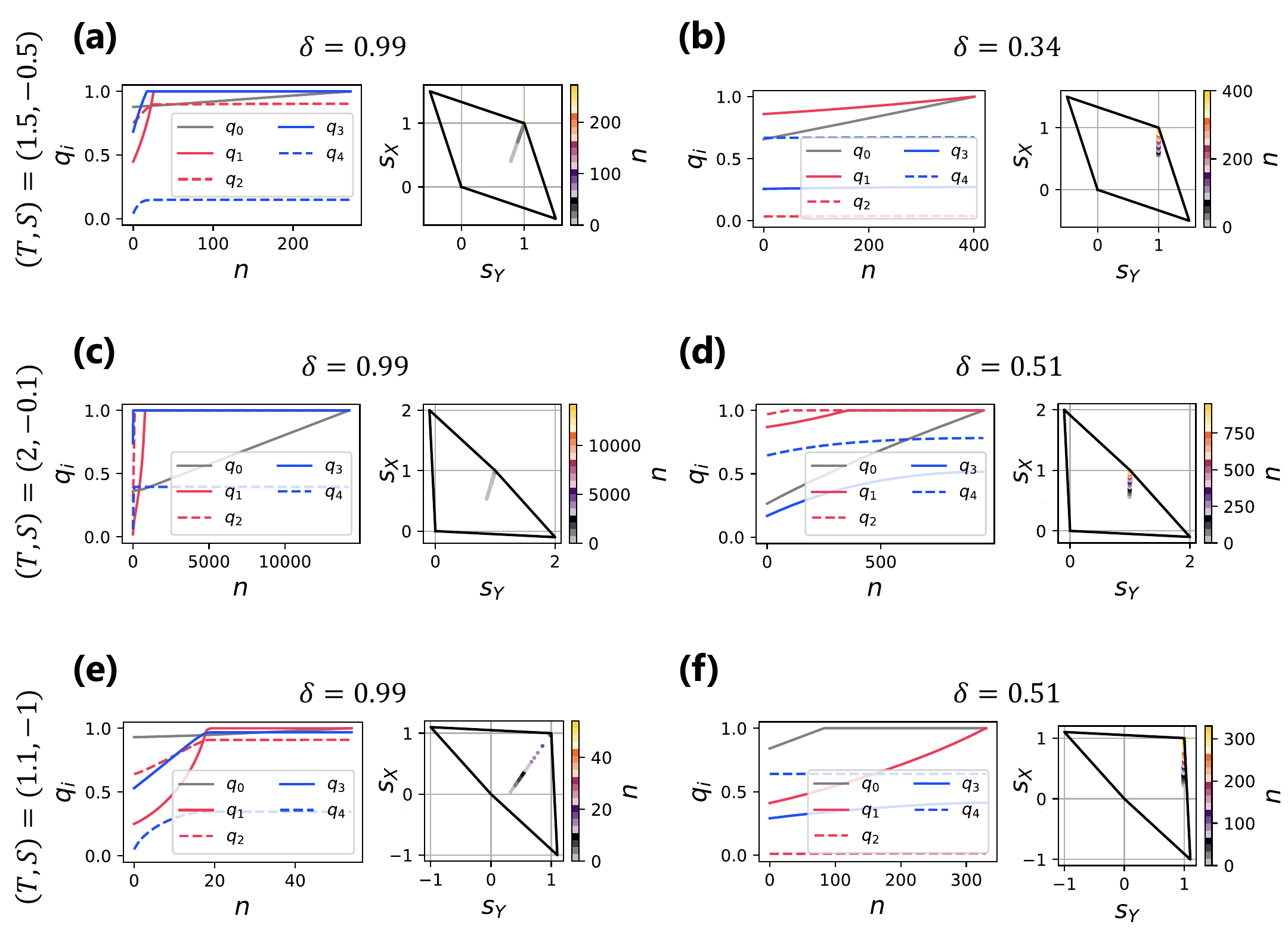}
\caption{
Improvement of $q_j$ (left) and $(s_Y,s_X)$ as time goes by, where 
\highlight{
{\bf (a)} $(T,S)=(1.5,-0.5),\ \delta=0.99$, $\bm q(0)=(0.877;0.449,0.751,0.684,0.04)$, and $\bm p=(1.0;1.0,0.5,0.8,0.3)$; %
{\bf (b)} $(T,S)=(1.5,-0.5),\ \delta=0.34$, $\bm q(0)=(0.657;0.86,0.035,0.257,0.668)$, and $\bm p=(1.0;1.0,0.0,1.0,0.0)$; %
{\bf (c)} $(T,S)=(2.0,-0.1),\ \delta=0.99$, $\bm q(0)=(0.36;0.02,0.28,0.74,0.06)$, and $\bm p=(1.0;1.0,0.3,0.726,0.35)$; %
{\bf (d)} $(T,S)=(2.0,-0.1),\ \delta=0.51$, $\bm q(0)=(0.265;0.869,0.970,0.168,0.645)$, and $\bm p=(1.0;1.0,0.0,0.135,0.0)$; %
{\bf (e)} $(T,S)=(1.1,-1.0),\ \delta=0.99$, $\bm q(0)=(0.930;0.250,0.640,0.530,0.050)$, and $\bm p=(1.0;1.0,0.45,0.749,0.1)$; %
and {\bf (f)} $(T,S)=(1.1,-1.0),\ \delta=0.51$, $\bm q(0)=(0.840;0.410,0.010,0.290,0.640)$, and $\bm p=(1.0;1.0,0.865,1.0,0.0)$. %
}
On the left side of each figure, the five lines show the changes in $\bm q$: gray line is $q_0$; red solid line is $q_1$; red dashed line is $q_2$; blue solid line is $q_3$; blue dashed line is $q_4$. 
On the right side, the color of the points indicates the change in time.  
}
\label{fig:T2_each_summary}
\end{figure}

\begin{figure}[btp]
\centering
\includegraphics[width=\columnwidth]{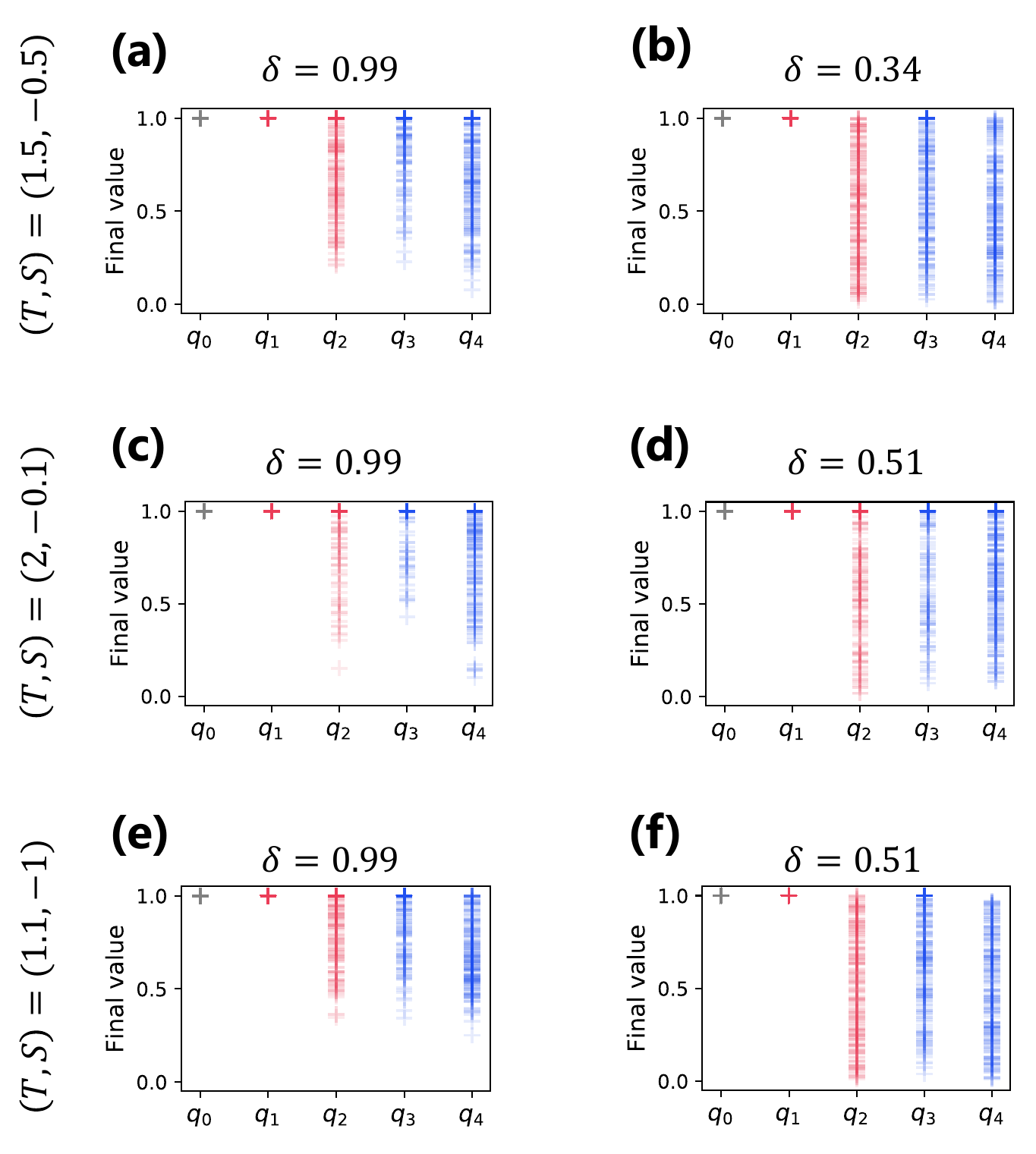}
\caption{
Final strategies obtained from the various initial strategies (which are 100 random instances), where %
\highlight{
{\bf (a)} $(T,S)=(1.5,-0.5),\ \delta=0.99$, and $\bm p=(1.0;1.0,0.5,0.8,0.3)$; %
{\bf (b)} $(T,S)=(1.5,-0.5),\ \delta=0.34$, and $\bm p=(1.0;1.0,0.0,1.0,0.0)$; %
{\bf (c)} $(T,S)=(2.0,-0.1),\ \delta=0.99$, and $\bm p=(1.0;1.0,0.3,0.726,0.35)$;
{\bf (d)} $(T,S)=(2.0,-0.1),\ \delta=0.51$, and $\bm p=(1.0;1.0,0.0,0.135,0.0)$;
{\bf (e)} $(T,S)=(1.1,-1.0),\ \delta=0.99$, and $\bm p=(1.0;1.0,0.45,0.749,0.1)$
and {\bf (f)} $(T,S)=(1.1,-1.0),\ \delta=0.51$, and $\bm p=(1.0;1.0,0.865,1.0,0.0)$. %
}
Each point shows the value of $q_j$ at the point where the simulation ends. %
}
\label{fig:T2_final_summary}
\end{figure}

Next, we also simulated the improvement of $\bm q$ using pcZD strategies, which satisfies $(p_0,p_1)=(1,1)$, verifying (T2). 
Similarly, we show the changes in $q_j$ and $(s_Y,s_X)$ in Fig.~\ref{fig:T2_each_summary}. 
%
We found that the updates are finished when $(q_0,q_1)=(1,1)$ is satisfied. 
Figure~\ref{fig:T2_final_summary} also shows the final strategies obtained from the various initial strategies, as indicated in Fig.~\ref{fig:finals}. 
%
From this figure, it is confirmed that $(q_0,q_1)=(1,1)$ is satisfied for every final strategy.

\section{Discussion}
In 2012, Press and Dyson found ZD strategies that have novel properties in the context of the RPD games. 
ZD strategies linearize the payoff relationship between themselves and their opponents. 
In addition, they predicted that pcZD strategies will lead an adaptive player to an unconditional cooperation from an unconditional defection. 
Chen and Zinger proved that the values of partial derivatives of the expected payoff expressed in the form of a determinant are positive (except for particular cases). 
Namely, each cooperation probability in the strategy vector increases, and any strategies will then reach an unconditional cooperation. 
However, it was still an open question whether the property exists even in the case of a discount factor. 
For this reason, we represented the expected payoff when considering the discount factor as the form of the determinants. 
Then, using a similar technique as in \cite{ChenZinger2014JTheorBiol}, we proved that pcZD strategies lead an adaptive player to an unconditional cooperation from any initial strategies. 
\highlight{The major difference from the related work~\cite{ChenZinger2014JTheorBiol} is that we focused on $p_0$ and $q_0$, which ~\cite{ChenZinger2014JTheorBiol} ignored but becomes necessary to consider due to the discount factor, and derived the conditions to reach an unconditional cooperation using them. }
Moreover, we numerically simulated how adaptive players improved their strategies, and then confirmed that each cooperation probability in their strategy vector basically increased. 
As a result, their strategies consistently reached an unconditional cooperation. 

We propose the following two research directions that are not addressed in this paper but are attractive nonetheless: 
First, introducing implementation or observation errors into the RPD games is a natural step. 
If the partial derivatives of the expected payoff with errors can be derived, we can calculate the destination of all adapting paths. 
Second, in this paper, we only proved the destinations of all adapting paths for the special case of a PD in which $T+S>P$. 
Thus, it is important to expand this payoff relationship to the general PD. 

\appendix
\setcounter{figure}{0}
\setcounter{equation}{0}
\renewcommand{\thefigure}{\Alph{section}.\arabic{figure}}
\renewcommand{\theequation}{\Alph{section}.\arabic{equation}}

\section{Condition that $(\bm I-\delta M)$ is inversible\label{app:inversible}}
The matrix $I-\delta M$ is regular and inversible when $0\le p_j\le1$, $0\le q_j\le1$, and $0<\delta<1$ as proven below: 
\begin{align}
  \det(I-\delta M) 
  &= \left|
  \begin{array}{rrrr}
	1-\delta p_1 q_1  &   -\delta p_1 (1-q_1)  &   -\delta (1-p_1) q_1  &   -\delta (1-p_1) (1-q_1)  \\
	 -\delta p_2 q_3  &  1-\delta p_2 (1-q_3)  &   -\delta (1-p_2) q_3  &   -\delta (1-p_2) (1-q_3)  \\
	 -\delta p_3 q_2  &   -\delta p_3 (1-q_2)  &  1-\delta (1-p_3) q_2  &   -\delta (1-p_3) (1-q_2)  \\
	 -\delta p_4 q_4  &   -\delta p_4 (1-q_4)  &   -\delta (1-p_4) q_4  &  1-\delta (1-p_4) (1-q_4) 
  \end{array}
  \right| \nonumber \\
  &= \left|
  \begin{array}{rrrr}
	1-\delta p_1 q_1  &  1-\delta p_1  &  1-\delta q_1  & -1-\delta +\delta p_1 +\delta q_1 \\
	 -\delta p_2 q_3  &  1-\delta p_2  &   -\delta q_3  &   -\delta +\delta p_2 +\delta q_3 \\
	 -\delta p_3 q_2  &   -\delta p_3  &  1-\delta q_2  &   -\delta +\delta p_3 +\delta q_2 \\
	 -\delta p_4 q_4  &   -\delta p_4  &   -\delta q_4  &  1-\delta +\delta p_4 +\delta q_4
  \end{array}
  \right| \nonumber \\
  &= \left|
  \begin{array}{rrrr}
	1-\delta p_1 q_1  &  1-\delta p_1  &  1-\delta q_1  &  1-\delta  \\
	 -\delta p_2 q_3  &  1-\delta p_2  &   -\delta q_3  &  1-\delta  \\
	 -\delta p_3 q_2  &   -\delta p_3  &  1-\delta q_2  &  1-\delta  \\
	 -\delta p_4 q_4  &   -\delta p_4  &   -\delta q_4  &  1-\delta
  \end{array}
  \right| \nonumber \\
  &= (1-\delta) \left|
  \begin{array}{rrrr}
	1-\delta p_1 q_1  &  1-\delta p_1  &  1-\delta q_1  &  1  \\
	 -\delta p_2 q_3  &  1-\delta p_2  &   -\delta q_3  &  1  \\
	 -\delta p_3 q_2  &   -\delta p_3  &  1-\delta q_2  &  1  \\
	 -\delta p_4 q_4  &   -\delta p_4  &   -\delta q_4  &  1
  \end{array}
  \right| \nonumber \\
  &= -(1-\delta) D(\bm p,\bm q,\bm 1_4) \nonumber
\end{align}

$\therefore$ $\det(I-\delta M)\ne0$ because we already obtained $\forall\bm p,\bm q\ D(\bm p,\bm q,\bm 1_4)>0$ when $0<\delta<1$ from Lemma 1.

\section{Factorization of the RHS of $\partial s_X/\partial q_i$\label{app:factorization}}
\subsection{Deriving partial derivatives of $s_X$ for $q_{\ell}\ (\ell\in\{1,2,3,4\})$ (Eq.~(\ref{eq:factorized}))}
\begin{figure}
\centering
\includegraphics[width=\columnwidth/3]{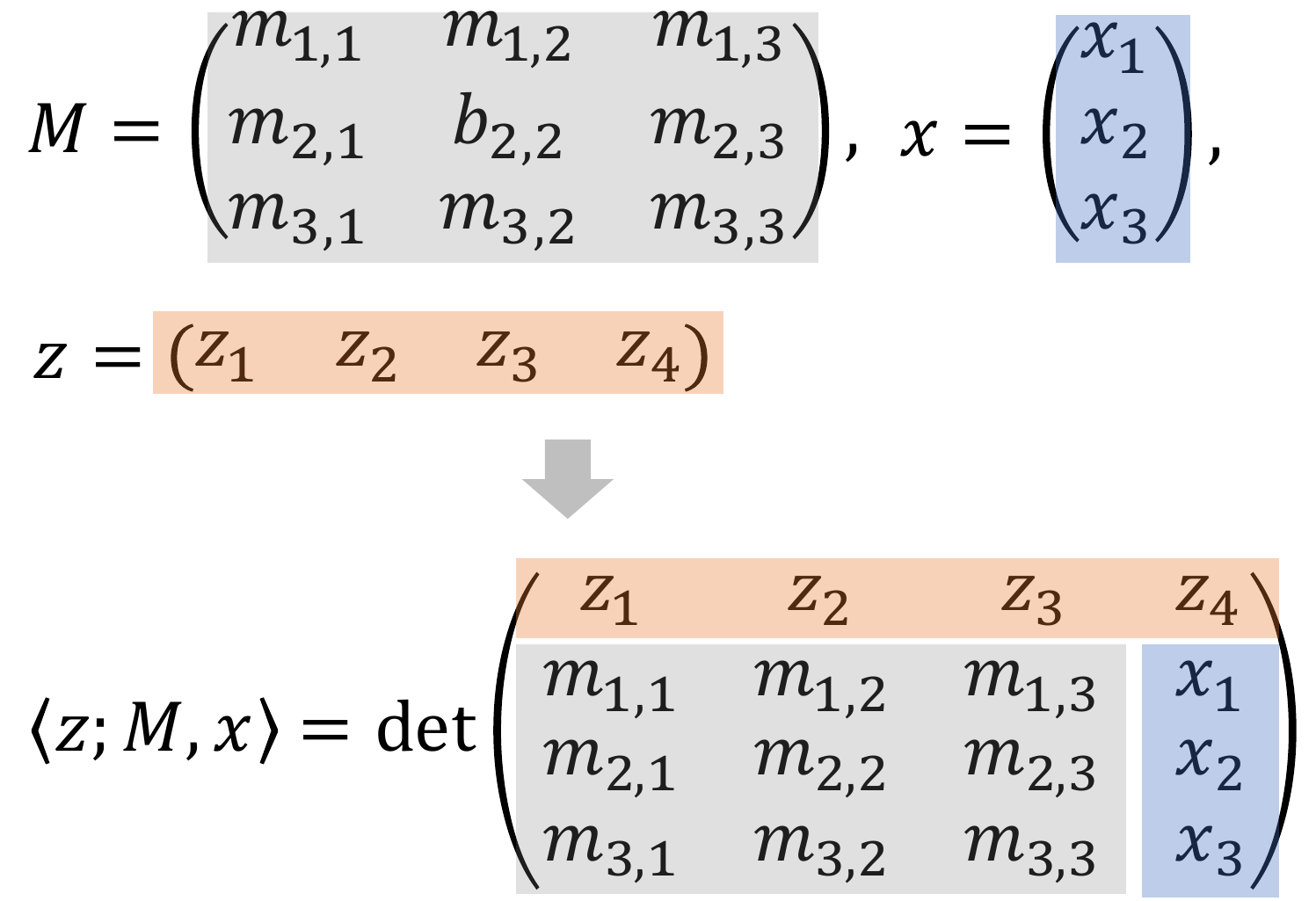}
\caption{
An $n\times n$ determinant $\left<\bm z;M,\bm x\right>$ (the $i$th row and $j$th column element is denoted as $d_{i,j}$) consists of $\bm z\in\mathbb{R}^4$ (orange), $M\in\mathbb{R}^{3\times3}$ (gray), $\bm x\in\mathbb{R}^3$ (blue), 
where $z_i$ is assigned to $d_{1,i}\ (i\in\{1,\cdots,n+1\})$, 
$x_j$ to $d_{j+1,n+1}\ (j\in\{1,\cdots,n\})$, 
and $m_{k,l}$ to $d_{k+1,l}\ (k,l\in\{1,\cdots,n\})$. 
}
\label{fig:det_zMx}
\end{figure}

We use Eq.~(20) of \cite{ChenZinger2014JTheorBiol} as follows: 
%
\begin{align}
\mathrm{det}	
 \left(
  \begin{array}{cc}
   \left<\bm z; M,\bm x\right>  &  \left<\tilde{\bm z}; M, \bm x'\right> \\
   & \\
   \left<\bm z'; M,\bm x\right>  &  \left<\bm z'; M, \bm x'\right>
  \end{array}
 \right)		
 &= \left<M\right>	
 \left(
  \sum_{1\le i\le j\le n} \left<M;\bm x, \bm x'\right>_{i,j} (z_i z'_j -z_j z'_i)
   + \sum_{1\le i\le n} \left<M; \tilde{z}_{n+1} \bm x -z_{n+1} \bm x'\right>_i z'_i
 \right), \label{eq:chen20}		
\end{align}
where $M\in\mathbb{R}^{n\times n}$ ($\left<M\right>$ denotes the determinant of $M$), 
$\bm x,\bm x'\in\mathbb{R}^n$, 
and $\bm z,\bm z',\tilde{\bm z}\in\mathbb{R}^{n+1}$. 
In addition, $\left<M;\bm x_1, \cdots, \bm x_k\right>_{i_1,\cdots,i_k}\ (k\le n)$ is the determinant of the matrix made of $M$ by replacing its columns numbered $i_1,\cdots,i_k$ by the column vectors $\bm x_1,\cdots,\bm x_k$ (illustrated in Fig.~\ref{fig:det_zMx}). 
%

We can regard $D(\bm p,\bm q,\bm f)$ and its partial derivatives in Eq.~(\ref{eq:partial}) as the style of $\left<\bm z; M, \bm x\right>$, where $n=3$, then apply Eq.~(\ref{eq:chen20}): 
%
\begin{align}
 D(\bm p, \bm q, \bm 1_4)^2 \frac{\partial s_X}{\partial q_{\ell}}	
 &= \mathrm{det}	
 \left(
  \begin{array}{cc}
   \left<\bm z_{\ell}; M_{\ell},\bm 1_3\right>  &  \left<\tilde{\bm z}_{\ell};M , \bm x'_{\ell}\right> \\
   & \\
   \delta \left<\bm z'_{\ell}; M_{\ell},\bm 1_3\right>  &  \delta \left<\bm z'_{\ell}; M, \bm x'_{\ell}\right>
  \end{array}
 \right), \label{eq:chenLHS}			
\end{align}
where
\begin{itemize}
\item $M_{\ell}$ is the $3\times3$ determinant, where $D(\bm p,\bm q,\bm f)$ is eliminated in the $\ell$th row and 4th column; 
\item $\bm z_{\ell}$ is the $\ell$th row vector of $D(\bm p,\bm q,\bm 1_4)$; 
\item $\bm z'_{\ell}$ is the $\ell$th row vector of $D(\bm p,\bm q,\bm f)$, which is partially differentiated for $q_{\ell}$ and then divided by $\delta$; 
\item $\tilde{\bm z}_{\ell}$ is the $\ell$th row vector of $D(\bm p,\bm q,\bm S_X)$ (the same as $\bm z$); and
\item $\bm x'_{\ell}$ is $\bm S_X$ with the $\ell$th element eliminated. 
\end{itemize}
Note that it is sufficient to partially differentiate only the $\ell$th row vector of $D(\bm p,\bm q,\bm f)$ for $\frac{\partial D(\bm p,\bm q,\bm f)}{\partial q_{\ell}}$ because $q_{\ell}$ does not exist in any other row. 
Thus, we obtain $\tilde{\bm z}_{\ell}=(p_{\lambda(\ell)}, 0, 1, 0)$, where
%
\begin{align*}
\lambda(\ell) \equiv \left\{
	\begin{array}{ll}
	\ell & ({\rm if}\ \ell\in\{1,4\}) \\
	5-\ell & ({\rm otherwise})
	\end{array}
\right.
.
\end{align*}

Transforming the RHS of Eq.~(\ref{eq:chenLHS}) using Eq.~(\ref{eq:chen20}), we obtain the following: 
%
%
\begin{align}
 D(\bm p, \bm q, \bm 1_4)^2 \frac{\partial s_X}{\partial q_{\ell}}	
 &= \delta \left<M_{\ell}\right>	
 \left(
  \sum_{i=1}^{2} \sum_{j=i+1}^{3} \left<M_{\ell};\bm 1_3, \bm x'_{\ell}\right>_{i,j} (z_{\ell; i} z'_{\ell; j} -z_{\ell; j} z'_{\ell; i})
   + \sum_{i=1}^{3} \left<M_{\ell}; \tilde{z}_{\ell;4} \bm 1_3 -\bm x'_{\ell}\right>_i z'_{\ell;i}
 \right), \label{eq:chenRHS}		
\end{align}
where $z_{\ell;i}$ denotes the $i$th element of $\bm z$. 

Next, we transform the last factor in Eq.~(\ref{eq:chenRHS}) (composed of two summation terms). 
Defining two summation terms as $S_1$ and $S_2$, we obtain the following expression: 
\begin{align}
 D(\bm p, \bm q, \bm 1_4)^2 \frac{\partial s_X}{\partial q_{\ell}}
 &= \delta \left<M_{\ell}\right> \left(S_1 +S_2\right). \label{eq:expand_sum}
\end{align}

Initially, we expand $S_1$ as follows: 
\begin{align}
 S_1 &\equiv \sum_{i=1}^{2} \sum_{j=i+1}^{3} \left<M_{\ell};\bm 1_3, \bm x'_{\ell}\right>_{i,j} (z_{\ell; i} z'_{\ell; j} -z_{\ell; j} z'_{\ell; i}) \nonumber \\
 &= \left|\bm 1_3\ \ \bm x'_{\ell}\ \ \bm m_{\ell;3}\right| (z_{\ell; 1} z'_{\ell; 2} -z_{\ell; 2} z'_{\ell; 1})
     + \left|\bm 1_3\ \ \bm m_{\ell;2}\ \ \bm x'_{\ell}\right| (z_{\ell; 1} z'_{\ell; 3} -z_{\ell; 3} z'_{\ell; 1}) \nonumber \\
 &\ \ + \left|\bm m_{\ell;1}\ \ \bm 1_3\ \ \bm x'_{\ell}\right| (z_{\ell; 2} z'_{\ell; 3} -z_{\ell; 3} z'_{\ell; 2}), \nonumber
\end{align}
where $\bm m_{\ell;i}$ denotes the $i$th column vector of $M_{\ell}$, and $\left|\bm a_1\ \ \bm a_2\ \ \cdots\ \ \bm a_n\right|$ denotes the $n\times n$ determinant composed of $\bm a_i\in\mathbb{R}^n$. 

Because each term has the common column vector $\bm 1_3$ and $\bm x'_{\ell}$, we can add these terms by replacing the columns of each determinant: 
\begin{align}
 S_1 &= \left|( Z_{2,3} \bm m_{\ell; 1} -Z_{1,3} \bm m_{\ell; 2} +Z_{1,2} \bm m_{\ell; 3})\ \ \bm 1_3\ \ \bm x'_{\ell}\right|, \nonumber
\end{align}
where $Z_{i,j}\equiv z_{\ell; i} z'_{\ell; j} -z_{\ell; j} z'_{\ell; i} $. 
Substituting $Z_{1,2}=-p_{\lambda(\ell)} z_{\ell; 2},\ Z_{1,3}=z_{\ell; 1}-p_{\lambda(\ell)} z_{\ell; 3},\ Z_{2,3}=z_{\ell; 2}$ obtained through $\bm z'_{\ell}=(p_{\lambda(\ell)}, 0, 1, 0)$, we obtain the following: 
\begin{align}
 S_1 &= \left|(z_{\ell; 2} \bm m_{\ell; 1} +(p_{\lambda(\ell)} z_{\ell; 3} -z_{\ell; 1}) \bm m_{\ell; 2} -p_{\lambda(\ell)} z_{\ell; 2} \bm m_{\ell; 3})\ \ \bm 1_3\ \ \bm x'_{\ell}\right|. \label{eq:S1}
\end{align}

Next, we transform $S_2$ as shown below: 
\begin{align}
 S_2 &\equiv \sum_{i=1}^{3} \left<M_{\ell}; \tilde{z}_{\ell;4} \bm 1_3 -\bm x'_{\ell}\right>_i z'_{\ell;i} \nonumber \\
 &= z'_{\ell; 1} \left|(\tilde{z}_{\ell; 4} \bm 1_3 -\bm x'_{\ell})\ \ \bm m_{\ell; 2}\ \ \bm m_{\ell;3}\right|
     + z'_{\ell; 2} \left|\bm m_{\ell; 1}\ \ (\tilde{z}_{\ell; 4} \bm 1_3 -\bm x'_{\ell})\ \ \bm m_{\ell; 3}\right| \nonumber \\
 &\ \ + z'_{\ell; 3} \left|\bm m_{\ell;1}\ \ \bm m_{\ell; 2}\ \ (\tilde{z}_{\ell; 4} \bm 1_3 -\bm x'_{\ell})\right|. \nonumber
\end{align}

Substituting $\bm z'_{\ell}=(p_{\lambda(\ell)}, 0, 1, 0)$, we obtain
\begin{align}
 S_2 
 &= p_{\lambda(\ell)} \left|(\tilde{z}_{\ell; 4} \bm 1_3 -\bm x'_{\ell})\ \ \bm m_{\ell; 2}\ \ \bm m_{\ell;3}\right|
   + \left|\bm m_{\ell;1}\ \ \bm m_{\ell; 2}\ \ (\tilde{z}_{\ell; 4} \bm 1_3 -\bm x'_{\ell})\right| \nonumber
 \\ %
 &= p_{\lambda(\ell)} \left|\bm m_{\ell;3}\ \ \bm m_{\ell; 2}\ \ (\bm x'_{\ell} -\tilde{z}_{\ell; 4} \bm 1_3)\right|
   -\left|\bm m_{\ell;1}\ \ \bm m_{\ell; 2}\ \ \bm x'_{\ell} -\tilde{z}_{\ell; 4} \bm 1_3\right| \nonumber
 \\ %
 &= \left|(p_{\lambda(\ell)} \bm m_{\ell;3} -\bm m_{\ell; 1})\ \ \bm m_{\ell; 2}\ \ (\bm x'_{\ell} -\tilde{z}_{\ell; 4} \bm 1_3)\right|. \label{eq:S2}
\end{align}

Therefore, we obtain the following expression by substituting Eq.~(\ref{eq:S1}) and (\ref{eq:S2}) for Eq.~(\ref{eq:expand_sum}): 
\begin{align}
 S_1 +S_2
 &= \left|(z_{\ell; 2} \bm m_{\ell; 1} +(p_{\lambda(\ell)} z_{\ell; 3} -z_{\ell; 1}) \bm m_{\ell; 2} -p_{\lambda(\ell)} z_{\ell; 2} \bm m_{\ell; 3})\ \ \bm 1_3\ \ \bm x'_{\ell}\right| \nonumber \\
 &\ \ \ \ +\left|(p_{\lambda(\ell)} \bm m_{\ell;3} -\bm m_{\ell; 1})\ \ \bm m_{\ell; 2}\ \ (\bm x'_{\ell} -\tilde{z}_{\ell; 4} \bm 1_3)\right| \nonumber 
 \\ \nonumber \\%
 &= \left|(z_{\ell; 2} \bm m_{\ell; 1} +(p_{\lambda(\ell)} z_{\ell; 3} -z_{\ell; 1}) \bm m_{\ell; 2} -p_{\lambda(\ell)} z_{\ell; 2} \bm m_{\ell; 3})\ \ \bm 1_3\ \ (\bm x'_{\ell} -\tilde{z}_{\ell; 4} \bm 1_3)\right| \nonumber \\
 &\ \ \ \ -\left|(\bm m_{\ell; 1} -p_{\lambda(\ell)} \bm m_{\ell;3})\ \ \bm m_{\ell; 2}\ \ (\bm x'_{\ell} -\tilde{z}_{\ell; 4} \bm 1_3)\right| \nonumber 
 \\ \nonumber \\%
 &= z_{\ell; 2} \left|(\bm m_{\ell; 1} -p_{\lambda(\ell)} \bm m_{\ell; 3})\ \ \bm 1_3\ \ (\bm x'_{\ell} -\tilde{z}_{\ell; 4} \bm 1_3)\right|
 +(p_{\lambda(\ell)} z_{\ell; 3} -z_{\ell; 1}) \left|\bm m_{\ell; 2}\ \ \bm 1_3\ \ (\bm x'_{\ell} -\tilde{z}_{\ell; 4} \bm 1_3)\right| \nonumber \\
 &\ \ \ \ -\left|(\bm m_{\ell; 1} -p_{\lambda(\ell)} \bm m_{\ell;3})\ \ \bm m_{\ell; 2}\ \ (\bm x'_{\ell} -\tilde{z}_{\ell; 4} \bm 1_3)\right| \nonumber 
 \\ \nonumber \\%
 &= z_{\ell; 2} \left|(\bm m_{\ell; 1} -p_{\lambda(\ell)} \bm m_{\ell; 3})\ \ \bm 1_3\ \ (\bm x'_{\ell} -\tilde{z}_{\ell; 4} \bm 1_3)\right| \nonumber \\
 &\ \ \ \ -\left|\left\{(p_{\lambda(\ell)} z_{\ell; 3} -z_{\ell; 1}) \bm 1_3 +\bm m_{\ell; 1} -p_{\lambda(\ell)} \bm m_{\ell; 3}\right\}\ \ \bm m_{\ell; 2}\ \ (\bm x'_{\ell} -\tilde{z}_{\ell; 4} \bm 1_3)\right| \nonumber
 \\ \nonumber \\%
 &= \left|\left\{(p_{\lambda(\ell)} z_{\ell; 3} -z_{\ell; 1}) \bm 1_3 +\bm m_{\ell; 1} -p_{\lambda(\ell)} \bm m_{\ell; 3}\right\}\ \ (z_{\ell; 2} \bm 1_3 -\bm m_{\ell; 2})\ \ (\bm x'_{\ell} -\tilde{z}_{\ell; 4} \bm 1_3)\right|. \label{eq:S1pS2}
\end{align}

Now, we define $\bm r_{\ell} \equiv (p_{\lambda(\ell)} z_{\ell; 3} -z_{\ell; 1}) \bm 1_3 +\bm m_{\ell; 1} -p_{\lambda(\ell)} \bm m_{\ell; 3}$ and $\bm r_{\ell}^T=(r_{\ell;1},r_{\ell;2},r_{\ell;3})$. 
Thus, 

\begin{align}
 S_1 +S_2
 &= \left|\bm r_{\ell}\ \ (z_{\ell; 2} \bm 1_3 -\bm m_{\ell; 2})\ \ (\bm x'_{\ell} -\tilde{z}_{\ell; 4} \bm 1_3)\right|. \label{eq:S1pS2_}
\end{align}

We then transform Eq.~(\ref{eq:S1pS2_}) further for each $\ell\in\{1,2,3,4\}$. 

\subsubsection{Case of $\ell=1$:}

Substituting 
$z_{1;2}=-1 +\delta p_1 +(1 -\delta) p_0$, 
$\tilde{z}_{1;4}=1$, 
$\bm m_{1;2}^T=(\delta p_3+(1-\delta)p_0, -1+\delta p_2+(1-\delta)p_0, \delta p_4+(1-\delta)p_0)$, 
and $\bm x'^T_1=(T,S,0)$ for Eq.~(\ref{eq:S1pS2_}), we obtain

\begin{align*}
\left|\bm r_1\ \ (z_{1;2}\bm 1_3-\bm m_{1;2})\ \ (\bm x'_1-\tilde{z}_{1;4}\bm 1_3)\right|
 &=\left|
  \begin{array}{ccc}
   r_{1; 1}  &  -1 +\delta p_1 -\delta p_3  &  T-1 \\
   r_{1; 2}  &  \delta p_1 -\delta p_2  &  S-1 \\
   r_{1; 3}  &  -1 +\delta p_1 -\delta p_4  &  -1
  \end{array}
 \right|. 
\end{align*}
%
Adding the third column multiplied by $(-1+\delta p_1-\delta p_4)$ to the second column, 
%
\begin{align*}
 &=\left|
  \begin{array}{ccc}
   r_{1; 1}  &  -1 +\delta p_1 -\delta p_3  -(T-1)(1-\delta p_1+\delta p_4) &  T-1 \\
   r_{1; 2}  &  \delta p_1 -\delta p_2 -(S-1)(1-\delta p_1+\delta p_4) &  S-1 \\
   r_{1; 3}  &  0  &  -1
  \end{array}
 \right|. 
\end{align*}
%
Adding the second row to the first row, 
%
\begin{align*}
 &=\left|
  \begin{array}{ccc}
   r_{1; 1}+r_{1; 2}  &  1 +2\delta p_4 -\delta p_2 -\delta p_3  -(T+S)(1-\delta p_1+\delta p_4) &  T+S-2 \\
   r_{1; 2}  &  \delta p_1 -\delta p_2 -(S-1)(1-\delta p_1+\delta p_4) &  S-1 \\
   r_{1; 3}  &  0  &  -1
  \end{array}
 \right|. 
\end{align*}
%
We then obtain the following by using Eq.~(\ref{eq:lincond}) for the $(2,2)$ element: 
%
\begin{align*}
 &=\left|
  \begin{array}{ccc}
   r_{1; 1}+r_{1; 2}  &  0 &  T+S-2 \\
   r_{1; 2}  &  \dot{p_2} -\dot{p_1}S +\delta p_4(1-S) &  S-1 \\
   r_{1; 3}  &  0  &  -1
  \end{array}
 \right|. 
\end{align*}
Thus, we obtain
%
\begin{align*}
 &=(\dot{p_2} -\dot{p_1}S +\delta p_4(1-S)) \left|
  \begin{array}{cc}
   r_{1; 1}+r_{1; 2}  &  T+S-2 \\
   r_{1; 3}  &  -1
  \end{array}
 \right|. 
\end{align*}
If we define 
\begin{align}
 \mathfrak{d_1} &\equiv
 \left|
  \begin{array}{cc}
   r_{1; 1}+r_{1; 2}  &  T+S-2 \\
   r_{1; 3}  &  -1
  \end{array}
 \right|, 
\end{align}
we obtain Eq.~(\ref{eq:factorized}) for $\ell=1$. 

\subsubsection{Case of $\ell=2$:}

Substituting 
$z_{2;2}=\delta p_3 +(1 -\delta) p_0$, 
$\tilde{z}_{2;4}=T$, 
$\bm m_{2;2}^T=(-1+\delta p_1+(1-\delta)p_0, -1+\delta p_2+(1-\delta)p_0, \delta p_4+(1-\delta)p_0)$, 
and $\bm x'^T_2=(1,S,0)$ for Eq.~(\ref{eq:S1pS2_}), we obtain

\begin{align*}
\left|\bm r_2\ \ (z_{2;2}\bm 1_3-\bm m_{2;2})\ \ (\bm x'_2-\tilde{z}_{2;4}\bm 1_3)\right|
 &=\left|
  \begin{array}{ccc}
   r_{2; 1}  &  1 +\delta p_3 -\delta p_1 &  1-T \\
   r_{2; 2}  &  1 +\delta p_3 -\delta p_2  &  S-T \\
   r_{2; 3}  &  \delta p_3 -\delta p_4  &  -T
  \end{array}
 \right|. 
\end{align*}
%
Subtracting the third row and the third row doubled from the first and second rows, respectively: 
%
\begin{align*}
 &=\left|
  \begin{array}{ccc}
   r_{2; 1}- r_{2; 3}  &  1 -\delta p_1 +\delta p_4 &  1 \\
   r_{2; 2}-2r_{2; 3}  &  1 +2\delta p_4 -\delta p_2 -\delta p_3  &  T+S \\
   r_{2; 3}  &  \delta p_3 -\delta p_4  &  -T
  \end{array}
 \right|. 
\end{align*}
%
Subtracting the third column multiplied by $1-\delta p_1+\delta p_4$ from the second column, 
%
\begin{align*}
 &=\left|
  \begin{array}{ccc}
   r_{2; 1}- r_{2; 3}  &  0  &  1 \\
   r_{2; 2}-2r_{2; 3}  &  1 +2\delta p_4 -\delta p_2 -\delta p_3 -(T+S)(1-\delta p_1 +\delta p_4)  &  T+S \\
   r_{2; 2}- r_{2; 3}  &  1 -\delta p_2 +\delta p_4 -S(1-\delta p_1+\delta p_4)  &  S
  \end{array}
 \right|. 
\end{align*}
%
We then obtain the following using Eq.~(\ref{eq:lincond}) for the (3,2) element: 
%
\begin{align*}
 &=\left|
  \begin{array}{ccc}
   r_{2; 1}- r_{2; 3}  &  0  &  1 \\
   r_{2; 2}-2r_{2; 3}  &  0  &  T+S \\
   r_{2; 2}- r_{2; 3}  &  \dot{p_2} -\dot{p_1}S +\delta p_4(1-S)  &  S
  \end{array}
 \right|. 
\end{align*}
Thus, we obtain
%
\begin{align*}
 &=(\dot{p_2} -\dot{p_1}S +\delta p_4(1-S)) \left|
  \begin{array}{cc}
   r_{2; 2}-2r_{2; 3}  &  T+S \\
   r_{2; 1}- r_{2; 3}  &  1
  \end{array}
 \right|. 
\end{align*}
If we define
\begin{align}
 \mathfrak{d_2} &\equiv
 \left|
  \begin{array}{cc}
   r_{2; 2}-2r_{2; 3}  &  T+S \\
   r_{2; 1}- r_{2; 3}  &  1
  \end{array}
 \right|, 
\end{align}
we obtain Eq.~(\ref{eq:factorized}) for $\ell=2$. 

\subsubsection{Case of $\ell=3$:}

Substituting 
$z_{3;2}=-1 +\delta p_2 +(1 -\delta) p_0$, 
$\tilde{z}_{3;4}=S$, 
$\bm m_{3;2}^T=(-1+\delta p_1+(1-\delta)p_0, \delta p_3+(1-\delta)p_0, \delta p_4+(1-\delta)p_0)$, 
and $\bm x'^T_3=(1,T,0)$ for Eq.~(\ref{eq:S1pS2_}), we obtain

\begin{align*}
\left|\bm r_3\ \ (z_{3;2}\bm 1_3-\bm m_{3;2})\ \ (\bm x'_3-\tilde{z}_{3;4}\bm 1_3)\right|
 &=\left|
  \begin{array}{ccc}
   r_{3; 1}  &      \delta p_2 -\delta p_1  &  1-S \\
   r_{3; 2}  &  -1 +\delta p_2 -\delta p_3  &  T-S \\
   r_{3; 3}  &  -1 +\delta p_2 -\delta p_4  &  -S
  \end{array}
 \right|. 
\end{align*}
%
Subtracting the third row and the third row doubled from the first and second rows, respectively, 
%
\begin{align*}
 &=\left|
  \begin{array}{ccc}
   r_{3; 1}- r_{3; 3}  &   1 -\delta p_1 +\delta p_4              &  1 \\
   r_{3; 2}-2r_{3; 3}  &   1+2\delta p_4 -\delta p_2 -\delta p_3  &  T+S \\
   r_{3; 3}            &  -1 +\delta p_2 -\delta p_4              &  -S
  \end{array}
 \right|. 
\end{align*}
%
Subtracting the third column multiplied by $1-\delta p_1+\delta p_4$ from the second column, 
%
\begin{align*}
 &=\left|
  \begin{array}{ccc}
   r_{3; 1}- r_{3; 3}  &  0  &  1 \\
   r_{3; 2}-2r_{3; 3}  &   1+2\delta p_4 -\delta p_2 -\delta p_3 -(T+S)(1-\delta p_1+\delta p_4)  &  T+S \\
   r_{3; 3}            &  -1 +\delta p_2 -\delta p_4 +S(1-\delta p_1+\delta p_4)  &  -S
  \end{array}
 \right|. 
\end{align*}
%
We then obtain the following using Eq.~(\ref{eq:lincond}) for the (3,2) element: 
%
\begin{align*}
 &=\left|
  \begin{array}{ccc}
   r_{3; 1}- r_{3; 3}  &  0  &  1 \\
   r_{3; 2}-2r_{3; 3}  &  0  &  T+S \\
   r_{3; 3}            &  -(\dot{p_2} -\dot{p_1}S +\delta p_4(1-S))  &  -S
  \end{array}
 \right|. 
\end{align*}
Thus, we obtain
%
\begin{align*}
 &=(\dot{p_2} -\dot{p_1}S +\delta p_4(1-S)) \left|
  \begin{array}{cc}
   r_{3; 1}- r_{3; 3}  &  1 \\
   r_{3; 2}-2r_{3; 3}  &  T+S
  \end{array}
 \right|. 
\end{align*}
If we define
\begin{align}
 \mathfrak{d_3} &\equiv
 \left|
  \begin{array}{cc}
   r_{3; 1}- r_{3; 3}  &  1 \\
   r_{3; 2}-2r_{3; 3}  &  T+S
  \end{array}
 \right|, 
\end{align}
we obtain Eq.~(\ref{eq:factorized}) for $\ell=3$. 

\subsubsection{Case of $\ell=4$:}

Substituting 
$z_{4;2}=\delta p_4 +(1 -\delta) p_0$, 
$\tilde{z}_{4;4}=0$, 
$\bm m_{4;2}^T=(-1+\delta p_1+(1-\delta)p_0, \delta p_3+(1-\delta)p_0, -1+\delta p_2+(1-\delta)p_0)$, 
and $\bm x'^T_4=(1,T,S)$ for Eq.~(\ref{eq:S1pS2_}), we obtain

\begin{align*}
\left|\bm r_4\ \ (z_{4;2}\bm 1_3-\bm m_{4;2})\ \ (\bm x'_4-\tilde{z}_{4;4}\bm 1_3)\right|
 &=\left|
  \begin{array}{ccc}
   r_{4; 1}  &  1 +\delta p_4 -\delta p_1  &  1 \\
   r_{4; 2}  &     \delta p_4 -\delta p_3  &  T \\
   r_{4; 3}  &  1 +\delta p_4 -\delta p_2  &  S
  \end{array}
 \right|. 
\end{align*}
%
Adding the third row to the second,  
%
\begin{align*}
 &=\left|
  \begin{array}{ccc}
   r_{4; 1}            &  1+ \delta p_4 -\delta p_1  &  1 \\
   r_{4; 2} +r_{4; 3}  &  1+2\delta p_4 -\delta p_2 -\delta p_3  &  T+S \\
   r_{4; 3}            &  1+ \delta p_4 -\delta p_2  &  S
  \end{array}
 \right|. 
\end{align*}
%
Subtracting the third column multiplied by $1-\delta p_1+\delta p_4$ from the second column, 
%
\begin{align*}
 &=\left|
  \begin{array}{ccc}
   r_{4; 1}            &  0  &  1 \\
   r_{4; 2} +r_{4; 3}  &  1+2\delta p_4 -\delta p_2 -\delta p_3 -(T+S)(1-\delta p_1+\delta p_4)  &  T+S \\
   r_{4; 3}            &  1+ \delta p_4 -\delta p_2 -S(1-\delta p_1+\delta p_4)  &  S
  \end{array}
 \right|. 
\end{align*}
%
We then obtain the following using Eq.~(\ref{eq:lincond}) for the (2,2) element:
%
\begin{align*}
 &=\left|
  \begin{array}{ccc}
   r_{4; 1}            &  0  &  1 \\
   r_{4; 2} +r_{4; 3}  &  0  &  T+S \\
   r_{4; 3}            &  \dot{p_2} -\dot{p_1}S +\delta p_4(1-S)  &  S
  \end{array}
 \right|. 
\end{align*}
Thus, we obtain
%
\begin{align*}
 &=(\dot{p_2} -\dot{p_1}S +\delta p_4(1-S)) \left|
  \begin{array}{cc}
   r_{4; 2} +r_{4; 3}  &  T+S \\
   r_{4; 1}            &  1
  \end{array}
 \right|. 
\end{align*}
If we define
\begin{align}
 \mathfrak{d_4} &\equiv
 \left|
  \begin{array}{cc}
   r_{4; 2} +r_{4; 3}  &  T+S \\
   r_{4; 1}            &  1
  \end{array}
 \right|, 
\end{align}
we obtain Eq.~(\ref{eq:factorized}) for $\ell=4$. 

\subsection{Deriving the partial derivative of $s_X$ for $q_0$ (Eq.~(\ref{eq:factorized_0}))}

By Eq.~(\ref{eq:Dpq1}), $\frac{\partial D(\bm p,\bm q,\bm 1_4)}{\partial q_0}=0$ because $D(\bm p,\bm q,\bm 1_4)$ does not contain $q_0$. 
Using this, we obtain
%
%
\begin{align}
 D(\bm p, \bm q, \bm 1_4)^2 \frac{\partial s_X}{\partial q_0} 
 &= \left|
  \begin{array}{cc}
   D(\bm p, \bm q, \bm 1_4)  &  D(\bm p, \bm q, \bm S_X)  \\
   \\
   \displaystyle \frac{\partial D(\bm p, \bm q, \bm 1_4)}{\partial q_0}  &  \displaystyle \frac{\partial D(\bm p, \bm q, \bm S_X)}{\partial q_0}
  \end{array}
 \right| \nonumber \\
 &= \left|
  \begin{array}{cc}
   D(\bm p, \bm q, \bm 1_4)  &  D(\bm p, \bm q, \bm S_X)  \\
   \\
   \displaystyle 0  &  \displaystyle \frac{\partial D(\bm p, \bm q, \bm S_X)}{\partial q_0}
  \end{array}
 \right|, \nonumber \\
 \nonumber \\
 \therefore D(\bm p,\bm q,\bm 1_4) \frac{\partial s_X}{\partial q_0}
 &= \frac{\partial D(\bm p, \bm q, \bm S_X)}{\partial q_0}. 
 \label{eq:partial_0}
\end{align}

Now, we derive the RHS of Eq.~(\ref{eq:partial_0}) through the following steps. 

Initially, each element of $D(\bm p,\bm q,\bm S_X)$ is as follows: 
%
\begin{align}
 D(\bm p, \bm q, \bm S_X)
 &= \left|
  \begin{array}{cccc}
   -1 +\delta p_1 q_1 +(1-\delta)p_0 q_0  &   -1 +\delta p_1 +(1-\delta)p_0   &   -1 +\delta q_1 +(1-\delta)q_0   &   1  \\
       \delta p_3 q_2 +(1-\delta)p_0 q_0  &       \delta p_3 +(1-\delta)p_0   &   -1 +\delta q_2 +(1-\delta)q_0   &   T  \\
       \delta p_2 q_3 +(1-\delta)p_0 q_0  &   -1 +\delta p_2 +(1-\delta)p_0   &       \delta q_3 +(1-\delta)q_0   &   S  \\
       \delta p_4 q_4 +(1-\delta)p_0 q_0  &       \delta p_4 +(1-\delta)p_0   &       \delta q_4 +(1-\delta)q_0   &   0
  \end{array}
 \right|. \nonumber
\end{align}
Subtracting the fourth row from the first, second, and third rows, respectively, we obtain
%
\begin{align}
 &= \left|
  \begin{array}{cccc}
   -1 +\delta p_1 q_1 -\delta p_4 q_4  &   -1 +\delta p_1 -\delta p_4   &   -1 +\delta q_1 -\delta q_4   &   1  \\
       \delta p_3 q_2 -\delta p_4 q_4  &       \delta p_3 -\delta p_4   &   -1 +\delta q_2 -\delta q_4   &   T  \\
       \delta p_2 q_3 -\delta p_4 q_4  &   -1 +\delta p_2 -\delta p_4   &       \delta q_3 -\delta q_4   &   S  \\
       \delta p_4 q_4 +(1-\delta)p_0 q_0  &       \delta p_4 +(1-\delta)p_0   &       \delta q_4 +(1-\delta)q_0   &   0
  \end{array}
 \right|. \nonumber
\end{align}
Thus, $\frac{\partial D(\bm p,\bm q,\bm S_X)}{\partial q_0}$ is obtained by partially differentiating the fourth row below: 
%
\begin{align*}
\frac{\partial D(\bm p,\bm q,\bm 1_4)}{\partial q_0}
 &= \left|
  \begin{array}{cccc}
   -1 +\delta p_1 q_1 -\delta p_4 q_4  &   -1 +\delta p_1 -\delta p_4   &   -1 +\delta q_1 -\delta q_4   &   1  \\
       \delta p_3 q_2 -\delta p_4 q_4  &       \delta p_3 -\delta p_4   &   -1 +\delta q_2 -\delta q_4   &   T  \\
       \delta p_2 q_3 -\delta p_4 q_4  &   -1 +\delta p_2 -\delta p_4   &       \delta q_3 -\delta q_4   &   S  \\
       (1-\delta)p_0  &   0   &   1-\delta   &   0
  \end{array}
 \right| \nonumber \\
 &= (1-\delta) \left|
  \begin{array}{cccc}
   -1 +\delta p_1 q_1 -\delta p_4 q_4  &   -1 +\delta p_1 -\delta p_4   &   -1 +\delta q_1 -\delta q_4   &   1  \\
       \delta p_3 q_2 -\delta p_4 q_4  &       \delta p_3 -\delta p_4   &   -1 +\delta q_2 -\delta q_4   &   T  \\
       \delta p_2 q_3 -\delta p_4 q_4  &   -1 +\delta p_2 -\delta p_4   &       \delta q_3 -\delta q_4   &   S  \\
       p_0  &   0   &   1   &   0
  \end{array}
 \right|. 
\end{align*}

Expanding the determinant using the (4,1) and (4,3) elements, 
\begin{align*}
 &\left|
  \begin{array}{cccc}
   -1 +\delta p_1 q_1 -\delta p_4 q_4  &   -1 +\delta p_1 -\delta p_4   &   -1 +\delta q_1 -\delta q_4   &   1  \\
       \delta p_3 q_2 -\delta p_4 q_4  &       \delta p_3 -\delta p_4   &   -1 +\delta q_2 -\delta q_4   &   T  \\
       \delta p_2 q_3 -\delta p_4 q_4  &   -1 +\delta p_2 -\delta p_4   &       \delta q_3 -\delta q_4   &   S  \\
       p_0  &   0   &   1   &   0
  \end{array}
 \right| \\
 &= \left(
	-p_0 \left|
	  \begin{array}{ccc}
	   -1 +\delta p_1 -\delta p_4   &   -1 +\delta q_1 -\delta q_4   &   1  \\
	       \delta p_3 -\delta p_4   &   -1 +\delta q_2 -\delta q_4   &   T  \\
	   -1 +\delta p_2 -\delta p_4   &       \delta q_3 -\delta q_4   &   S
	  \end{array}
	\right|
	-\left|
	  \begin{array}{ccc}
	   -1 +\delta p_1 q_1 -\delta p_4 q_4  &   -1 +\delta p_1 -\delta p_4   &   1  \\
		   \delta p_3 q_2 -\delta p_4 q_4  &       \delta p_3 -\delta p_4   &   T  \\
		   \delta p_2 q_3 -\delta p_4 q_4  &   -1 +\delta p_2 -\delta p_4   &   S
	  \end{array}
	\right|
\right). 
\end{align*}
Combining both determinants by replacing the first and second columns of the former determinant, we obtain
%
\begin{align*}
 &= \left|
  \begin{array}{ccc}
   (-1 +\delta q_1 -\delta q_4) p_0 -(-1 +\delta p_1 q_1 -\delta p_4 q_4)  &   -1 +\delta p_1 -\delta p_4   &   1  \\
   (-1 +\delta q_2 -\delta q_4) p_0 -(    \delta p_3 q_2 -\delta p_4 q_4)  &       \delta p_3 -\delta p_4   &   T  \\
   (    \delta q_3 -\delta q_4) p_0 -(    \delta p_2 q_3 -\delta p_4 q_4)  &   -1 +\delta p_2 -\delta p_4   &   S
  \end{array}
\right|. 
\end{align*}
%
Now, if we define the first column as $\bm u^T=(u_1, u_2, u_3)$, we can then rewrite it as
%
\begin{align*}
 &= \left|
  \begin{array}{ccc}
   u_1  &   -1 +\delta p_1 -\delta p_4   &   1  \\
   u_2  &       \delta p_3 -\delta p_4   &   T  \\
   u_3  &   -1 +\delta p_2 -\delta p_4   &   S
  \end{array}
\right|. 
\end{align*}
Adding the third row to the second, 
%
\begin{align*}
 &= \left|
  \begin{array}{ccc}
   u_1  &   -1 +\delta p_1 -\delta p_4   &   1  \\
   u_2+u_3  &  -1 +\delta p_2 +\delta p_3 -2\delta p_4   &   T+S  \\
   u_3  &   -1 +\delta p_2 -\delta p_4   &   S
  \end{array}
\right|. 
\end{align*}
Adding the third column multiplied by $1-\delta p_1+\delta p_4$ to the second column, 
%
\begin{align*}
 &= \left|
  \begin{array}{ccc}
   u_1  &   0   &   1  \\
   u_2+u_3  &  -1 +\delta p_2 +\delta p_3 -2\delta p_4 +(T+S)(1-\delta p_1+\delta p_4)  &   T+S  \\
   u_3  &   -1 +\delta p_2 -\delta p_4 +S(1-\delta p_1+\delta p_4)  &   S
  \end{array}
\right|. 
\end{align*}
%
We then obtain the following using Eq.~(\ref{eq:lincond}) for the (2,2) element: 
%
\begin{align*}
 &= \left|
  \begin{array}{ccc}
   u_1  &   0   &   1  \\
   u_2+u_3  &  0  &   T+S  \\
   u_3  &   -(\dot{p_2}-\dot{p_1}S+\delta p_4(1-S))  &   S
  \end{array}
\right|. 
\end{align*}
Thus, we obtain
%
\begin{align*}
 &= (\dot{p_2}-\dot{p_1}S+\delta p_4(1-S)) \left|
  \begin{array}{ccc}
   u_1      &  1   \\
   u_2+u_3  &  T+S \\
  \end{array}
\right|. 
\end{align*}
If we define
%
\begin{align}
 \mathfrak{d}_0 &\equiv \left|
  \begin{array}{ccc}
   u_1      &  1   \\
   u_2+u_3  &  T+S \\
  \end{array}
\right|, 
\end{align}
%
we obtain Eq.~(\ref{eq:factorized_0}).

\section{Conditions for $\partial s_Y/\partial q_{\ell}=0\ (\ell\in\{1,2,3,4\})$\label{app:partial1234_conds}}
We show the conditions for $\partial s_Y/\partial q_{\ell}=0\ (\ell\in\{1,2,3,4\})$ below: 

\begin{itemize}
 \item $\partial s_Y/\partial q_1=0$ if and only if
 \begin{itemize}
  \item $(q_0, q_3, q_4) = (0, 0, 0)$ or
  \item $(p_0, p_4, q_0, q_4) = (0, 0, 0, 0)$ or
  \item $(p_0, q_2, q_3, q_4) = (0, 0, 0, 0)$ or
  \item $(p_2, q_0, q_2, q_4) = (0, 0, 0, 0)$ or
  \item $(p_4, q_0, q_2, q_3) = (0, 0, 0, 0)$ or
  \item $(p_0, p_2, q_0, q_2, q_4) = (0, 0, 1, 0, 0)$ or
  \item $(p_0, p_4, q_0, q_2, q_3) = (0, 0, 1, 0, 0)$ or
  \item $(p_2, p_4, q_0, q_2, q_4) = (0, 0, 0, 0, 1)$ or
  \item $(p_0, p_2, p_4, q_0, q_2, q_4) = (0, 0, 0, 1, 0, 1)$ or
  \item $(p_0, p_2, p_3, q_0, q_2, q_3) = (1, 0, 1, 0, 0, 1)$ or
  \item $(p_0, p_2, p_3, q_0, q_2, q_3) = (0, 0, 1, 1, 0, 1)$.
 \end{itemize}
 \item $\partial s_Y/\partial q_2=0$ if and only if
 \begin{itemize}
  \item $(q_0, q_3, q_4) = (0, 0, 0)$ or 
  \item $(p_0, p_1, q_0, q_1) = (1, 1, 1, 1)$ or
  \item $(p_0, p_4, q_0, q_4) = (0, 0, 0, 0)$ or
  \item $(p_0, q_1, q_3, q_4) = (1, 0, 0, 0)$. 
 \end{itemize}
 \item $\partial s_Y/\partial q_3=0$ if and only if
 \begin{itemize}
  \item $(q_0, q_1, q_2) = (1, 1, 1)$ or
  \item $(p_0, p_1, q_0, q_1) = (1, 1, 1, 1)$ or 
  \item $(p_0, p_4, q_0, q_4) = (0, 0, 0, 0)$ or 
  \item $(p_0, q_1, q_2, q_4) = (0, 1, 1, 1)$ or
  \item $(p_0, p_4, q_0, q_1, q_2) = (0, 0, 0, 0, 0)$. 
 \end{itemize}
 \item $\partial s_Y/\partial q_4=0$ if and only if
 \begin{itemize}
  \item $(q_0, q_1, q_2) = (1, 1, 1)$ or
  \item $(p_0, p_1, q_0, q_1) = (1, 1, 1, 1)$ or
  \item $(p_0, q_1, q_2, q_3) = (1, 1, 1, 1)$ or
  \item $(p_1, q_0, q_2, q_3) = (1, 1, 1, 1)$ or
  \item $(p_3, q_0, q_1, q_3) = (1, 1, 1, 1)$ or
  \item $(p_0, p_3, q_0, q_1, q_3) = (1, 1, 0, 1, 1)$ or
  \item $(p_1, p_3, q_0, q_1, q_3) = (1, 1, 1, 0, 1)$ or
  \item $(p_0, p_1, p_3, q_1, q_2, q_3) = (1, 1, 1, 0, 1, 1)$ or
  \item $(p_0, p_2, p_3, q_0, q_2, q_3) = (1, 0, 1, 0, 1, 1)$. 
 \end{itemize}
\end{itemize}

\section{Confirmation of increasing $\nu$\label{app:numcal}}
\setcounter{figure}{0}
\setcounter{equation}{0}
\highlight{
We simulated the global adaptions by increasing learning rate $\nu$ that controls the step size of numerical calculations as shown in Fig.~{\ref{fig:global}}. 
It shows that all adapting paths reach the final state even if the learning rate is high, thereby Theorem~{\ref{theor:main}} is robust.
}
\begin{figure}[hbtp]
\centering
\includegraphics[width=\textwidth-20mm]{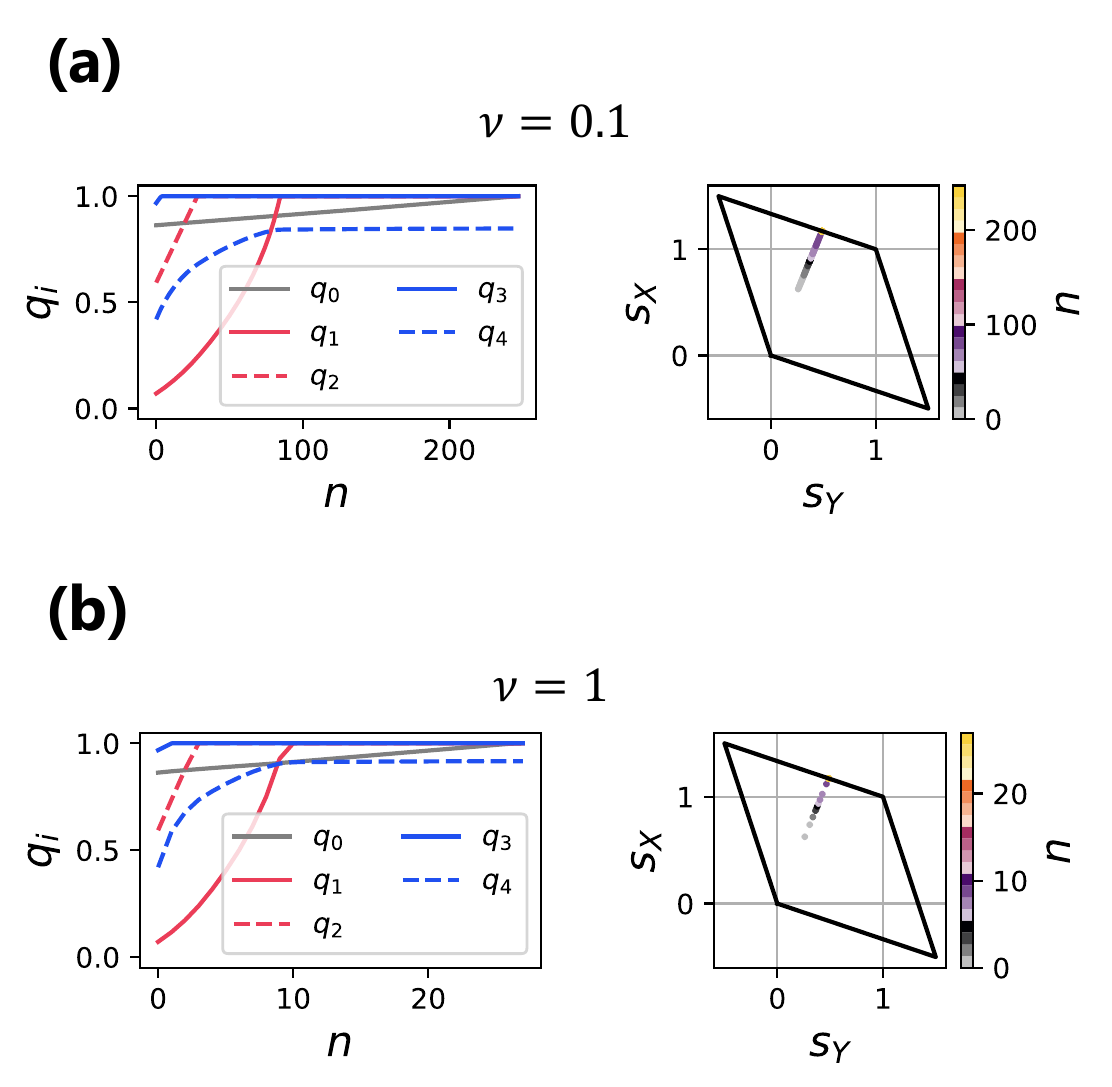}
\caption{
Improvement of $q_j$ (left) and $(s_Y,s_X)$ as time goes by where $\nu=1$, $\delta=0.99,\ (T,S)=(1.5,-0.5)$, $\bm q_0=(0.863;0.071,0.593,0.963,0.420)$, and $\bm p=(0.0; 0.75, 0.25, 0.5, 0.0)$. 
In the left figure, the five lines show the changes in $\bm q$: gray line is $q_0$; red solid line is $q_1$; red dashed line is $q_2$; blue solid line is $q_3$; blue dashed line is $q_4$. 
In the right figure, the color of the points indicates the change in time.  
}
\label{fig:global}
\end{figure}

\section*{Acknowledgment}
This study was partly supported by JSPS KAKENHI, Grant Number JP19K04903 (G.I.).

\section*{Additional Information}
Declarations of interest: none. 


\begin{thebibliography}{10}

\bibitem{Trivers1971QRevBiol}
R.~L. Trivers.
\newblock The evolution of reciprocal altruism.
\newblock {\em Q. Rev. Biol.}, 46:35--57, 1971.

\bibitem{Nowak2006book}
M.~A. Nowak.
\newblock {\em Evolutionary Dynamics}.
\newblock Harvard University Press, Cambridge, MA, 2006.

\bibitem{Sigmund2010book}
K.~Sigmund.
\newblock {\em The Calculus of Selfishness}.
\newblock Princeton University Press, Princeton, NJ, 2010.

\bibitem{Akin2016ErgodicTheory}
E.~Akin.
\newblock The iterated prisoner's dilemma: Good strategies and their dynamics.
\newblock {\em Ergodic Theory, Advances in Dynamics}, (de Gruyter, Berlin,
  2016):77--107, 2016.

\bibitem{Adami2013NatComm}
C.~Adami and A.~Hintze.
\newblock {Evolutionary instability of zero-determinant strategies demonstrates
  that winning is not everything}.
\newblock {\em Nat. Comm.}, 4:2193, 2013.

\bibitem{Hilbe2013PNAS}
C.~Hilbe, M.~A. Nowak, and K.~Sigmund.
\newblock {Evolution of extortion in Iterated Prisoner's Dilemma games}.
\newblock {\em Proc. Natl. Acad. Sci. USA}, 110:6913--6918, 2013.

\bibitem{Hilbe2013PlosOne-zd}
C.~Hilbe, M.~A. Nowak, and A.~Traulsen.
\newblock {Adaptive dynamics of extortion and compliance}.
\newblock {\em PLoS ONE}, 8:e77886, 2013.

\bibitem{ChenZinger2014JTheorBiol}
J.~Chen and A.~Zinger.
\newblock {The robustness of zero-determinant strategies in Iterated Prisoner's
  Dilemma games}.
\newblock {\em J. Theor. Biol.}, 357:46--54, 2014.

\bibitem{Szolnoki2014PhysRevE-zd}
A.~Szolnoki and M.~Perc.
\newblock {Evolution of extortion in structured populations}.
\newblock {\em Phys. Rev. E}, 89:022804, 2014.

\bibitem{Szolnoki2014SciRep-zd}
A.~Szolnoki and M.~Perc.
\newblock {Defection and extortion as unexpected catalysts of unconditional
  cooperation in structured populations}.
\newblock {\em Sci. Rep.}, 4:5496, 2014.

\bibitem{WuRong2014PhysRevE-zd}
Z.~X. Wu and Z.~Rong.
\newblock {Boosting cooperation by involving extortion in spatial prisoner's
  dilemma games}.
\newblock {\em Phys. Rev. E}, 90:062102, 2014.

\bibitem{Hilbe2015JTheorBiol}
C.~Hilbe, B.~Wu, A.~Traulsen, and M.~A. Nowak.
\newblock {Evolutionary performance of zero-determinant strategies in
  multiplayer games}.
\newblock {\em J. Theor. Biol.}, 374:115--124, 2015.

\bibitem{LiuLi2015PhysicaA-zd}
J.~Liu, Y.~Li, C.~Xu, and P.~M. Hui.
\newblock {Evolutionary behavior of generalized zero-determinant strategies in
  iterated prisoner's dilemma}.
\newblock {\em Physica A}, 430:81--92, 2015.

\bibitem{Xu2017PhysRevE-zd}
X.~Xu, Z.~Rong, Z.~X. Wu, T.~Zhou, and C.~K. Tse.
\newblock {Extortion provides alternative routes to the evolution of
  cooperation in structured populations}.
\newblock {\em Phys. Rev. E}, 95:052302, 2017.

\bibitem{Wang2019Chaos}
J.~Wang and J.~Guo.
\newblock {A synergy of punishment and extortion in cooperation dilemmas driven
  by the leader}.
\newblock {\em Chaos Solitons Fractals}, 119:263--268, 2019.

\bibitem{Stewart2013PNAS}
A.~J. Stewart and J.~B. Plotkin.
\newblock {From extortion to generosity, evolution in the Iterated Prisoner's
  Dilemma}.
\newblock {\em Proc. Natl. Acad. Sci. USA}, 110:15348--15353, 2013.

\bibitem{Mao2018EPL}
Y.~Mao, X.~Xu, Z.~Rong, and Z.~X. Wu.
\newblock {The emergence of cooperation-extortion alliance on scale-free
  networks with normalized payoff}.
\newblock {\em EPL}, 122:50005, 2018.

\bibitem{Xu2019Neurocomputing}
X.~Xu, Z.~Rong, Z.~Tian, and Z.~X. Wu.
\newblock {Timescale diversity facilitates the emergence of
  cooperation-extortion alliances in networked systems}.
\newblock {\em Neurocomputing}, 350:195--201, 2019.

\bibitem{Hao2015PhysRevE}
D.~Hao, Z.~Rong, and T.~Zhou.
\newblock Extortion under uncertainty: Zero-determinant strategies in noisy
  games.
\newblock {\em Phys. Rev. E}, 91:052803, 2015.

\bibitem{MamiyaIchinose2019JTheorBiol}
A.~Mamiya and G.~Ichinose.
\newblock Strategies that enforce linear payoff relationships under observation
  errors in repeated prisoner's dilemma game.
\newblock {\em J. Theor. Biol.}, 477:63--76, 2019.

\bibitem{Hilbe2014PNAS-zd}
C.~Hilbe, B.~Wu, A.~Traulsen, and M.~A. Nowak.
\newblock {Cooperation and control in multiplayer social dilemmas}.
\newblock {\em Proc. Natl. Acad. Sci. USA}, 111:16425--16430, 2014.

\bibitem{Pan2015SciRep-zd}
L.~Pan, D.~Hao, Z.~Rong, and T.~Zhou.
\newblock {Zero-determinant strategies in iterated public goods game}.
\newblock {\em Sci. Rep.}, 5:13096, 2015.

\bibitem{Milinski2016NatComm}
M.~Milinski, C.~Hilbe, D.~Semmann, R.~Sommerfeld, and J.~Marotzke.
\newblock {Humans choose representatives who enforce cooperation in social
  dilemmas through extortion}.
\newblock {\em Nat. Comm.}, 7:10915, 2016.

\bibitem{Stewart2016PNAS}
A.~J. Stewart, T.~L. Parsons, and J.~B. Plotkin.
\newblock {Evolutionary consequences of behavioral diversity}.
\newblock {\em Proc. Natl. Acad. Sci. USA}, 113:E7003--E7009, 2016.

\bibitem{UedaTanaka2020PlosOne}
M.~Ueda and T.~Tanaka.
\newblock Linear algebraic structure of zero-determinant strategies in repeated
  games.
\newblock {\em PLoS ONE}, 15:e0230973, 2020.

\bibitem{Mcavoy2016PNAS}
A.~{McAvoy} and C.~Hauert.
\newblock {Autocratic strategies for iterated games with arbitrary action
  spaces}.
\newblock {\em Proc. Natl. Acad. Sci. USA}, 113:3573--3578, 2016.

\bibitem{Mcavoy2017TheorPopulBiol}
A.~Mc{A}voy and C.~Hauert.
\newblock {Autocratic strategies for alternating games}.
\newblock {\em Theor. Popul. Biol.}, 113:13--22, 2017.

\bibitem{EngelFeigel2019ApplMathComput}
M.~A. Taha and A.~Ghoneim.
\newblock Zero-determinant strategies in repeated asymmetric games.
\newblock {\em Appl. Math. Comput.}, 369:124862, 2020.

\bibitem{EngelFeigel2018PhysRevE}
A.~Engel and A.~Feigel.
\newblock Single equalizer strategy with no information transfer for conflict
  escalation.
\newblock {\em Phys. Rev. E}, 98:012415, 2018.

\bibitem{Hilbe2014NatComm}
C.~Hilbe, T.~R{\"{o}}hl, and M.~Milinski.
\newblock {Extortion subdues human players but is finally punished in the
  prisoner's dilemma}.
\newblock {\em Nat. Comm.}, 5:3976, 2014.

\bibitem{Wang2016NatComm-zd}
Z.~Wang, Y.~Zhou, J.~W. Lien, J.~Zheng, and B.~Xu.
\newblock {Extortion can outperform generosity in the iterated prisoner's
  dilemma}.
\newblock {\em Nat. Comm.}, 7:11125, 2016.

\bibitem{Hilbe2016PlosOne}
C.~Hilbe, K.~Hagel, and M.~Milinski.
\newblock {Asymmetric power boosts extortion in an economic experiment}.
\newblock {\em PLoS ONE}, 11:e0163867, 2016.

\bibitem{Becks2019NatComm}
L.~Becks and M.~Milinski.
\newblock {Extortion strategies resist disciplining when higher competitiveness
  is rewarded with extra gain}.
\newblock {\em Nat. Comm.}, 10:783, 2019.

\bibitem{Press2012PNAS}
W.~H. Press and F.~J. Dyson.
\newblock {Iterated Prisoner's Dilemma contains strategies that dominate any
  evolutionary opponent}.
\newblock {\em Proc. Natl. Acad. Sci. USA}, 109:10409--10413, 2012.

\bibitem{Hilbe2015GamesEconBehav}
C.~Hilbe, A.~Traulsen, and K.~Sigmund.
\newblock {Partners or rivals? Strategies for the iterated prisoner's dilemma}.
\newblock {\em Games Econ. Behav.}, 92:41--52, 2015.

\bibitem{IchinoseMasuda2018JTheorBiol}
G.~Ichinose and N.~Masuda.
\newblock {Zero-determinant strategies in finitely repeated games}.
\newblock {\em J. Theor. Biol.}, 438:61--77, 2018.

\bibitem{GovaertCao2019arXiv}
A.~Govaert and M.~Cao.
\newblock Zero-determinant strategies in finitely repeated n-player games.
\newblock {\em arXiv}, 2019.

\bibitem{MamiyaIchinose2021JTheorBiol}
Azumi Mamiya, Daiki Miyagawa, and Genki Ichinose.
\newblock Conditions for the existence of zero-determinant strategies under
  observation errors in repeated games.
\newblock {\em Journal of Theoretical Biology}, 526:110810, 2021.

\end{thebibliography}
\end{document}